\documentclass[8pt,twoside,a4paper]{article}

\newcommand{\mytitle}{A Quantitative Model for Optical Coherence Tomography}
\title{\mytitle}

\date{November 26, 2019}
\author{Leopold Veselka$^1$\\{\footnotesize\href{mailto:lepold.veselka@univie.ac.at}{leopold.veselka@univie.ac.at}}
\and Lisa Krainz$^2$\\{\footnotesize\href{mailto:lisa.krainz@meduniwien.ac.at}{lisa.krainz@meduniwien.ac.at}}
\and Leonidas Mindrinos$^1$\\{\footnotesize\href{mailto:leonidas.mindrinos@univie.ac.at}{leonidas.mindrinos@univie.ac.at}}
\and Wolfgang Drexler$^1$\\{\footnotesize\href{mailto:wolfgang.drexler@meduniwien.ac.at}{wolfgang.drexler@meduniwien.ac.at}}
\and Peter Elbau$^2$\\{\footnotesize\href{mailto:peter.elbau@univie.ac.at}{peter.elbau@univie.ac.at}}}

\usepackage[utf8]{inputenc}

\usepackage[T1]{fontenc}
\usepackage{lmodern}

\usepackage{amsmath}
\numberwithin{equation}{section}

\usepackage{amssymb}

\usepackage{microtype}

\usepackage[]{algorithm2e}

\usepackage[backend=bibtex,maxnames=15]{biblatex}
\usepackage{graphicx}
\graphicspath{{images/}}

\usepackage{tikz}
\usepackage{pgfplots}
\pgfplotsset{compat=newest}

\title{A Quantitative Model for Optical Coherence Tomography}
\author{Leopold Veselka$^1$\\{\footnotesize\href{mailto:lepold.veselka@univie.ac.at}{leopold.veselka@univie.ac.at}}
\and Lisa Krainz$^2$\\{\footnotesize\href{mailto:lisa.krainz@meduniwien.ac.at}{lisa.krainz@meduniwien.ac.at}}
\and Leonidas Mindrinos$^1$\\{\footnotesize\href{mailto:leonidas.mindrinos@univie.ac.at}{leonidas.mindrinos@univie.ac.at}}
\and Wolfgang Drexler$^2$\\{\footnotesize\href{mailto:wolfgang.drexler@meduniwien.ac.at}{wolfgang.drexler@meduniwien.ac.at}}
\and Peter Elbau$^1$\\{\footnotesize\href{mailto:peter.elbau@univie.ac.at}{peter.elbau@univie.ac.at}}}
\date{}

\usepackage[pdftex,colorlinks=true,linkcolor=blue,citecolor=green,urlcolor=blue,bookmarks=true,bookmarksnumbered=true]{hyperref}
\hypersetup
{
    pdfauthor={Leopold Veselka, Lisa Krainz, Leonidas Mindrinos, Wolfgang Drexler, Peter Elbau},
    pdfsubject={},
    pdftitle={\mytitle},
    pdfkeywords={}
}

\usepackage[hyperref,amsmath,thmmarks]{ntheorem}
\usepackage{aliascnt}

\newtheorem{lemma}{Lemma}[section]

\newaliascnt{proposition}{lemma}

\aliascntresetthe{proposition}

\newaliascnt{corollary}{lemma}

\aliascntresetthe{corollary}

\newaliascnt{theorem}{lemma}
\newtheorem{theorem}[theorem]{Theorem}
\aliascntresetthe{theorem}

\theorembodyfont{\normalfont}
\newaliascnt{definition}{lemma}

\aliascntresetthe{definition}

\newaliascnt{assumption}{lemma}

\aliascntresetthe{assumption}

\newaliascnt{example}{lemma}

\aliascntresetthe{example}

\theoremstyle{nonumberplain}
\theoremseparator{:}
\theoremheaderfont{\normalfont\itshape}

\theoremsymbol{\ensuremath{\square}}
\newtheorem{proof}{Proof}

\usepackage[a4paper,centering,bindingoffset=0cm,marginpar=2cm,margin=2.5cm]{geometry}

\usepackage[pagestyles]{titlesec}
\usepackage[font=footnotesize,format=plain,labelfont=sc,textfont=sl,width=0.75\textwidth,labelsep=period]{caption}
\titleformat{\section}[block]{\large\sc\filcenter}{\thesection.}{0.5ex}{}[]
\titleformat{\subsection}[runin]{\bf}{\thesubsection.}{0.5ex}{}[.]

\newpagestyle{headers}%
{%
	\headrule
	\sethead%
		[\footnotesize\thepage]%
		[\footnotesize\sc L. Veselka, L.Krainz, L. Mindrinos, W. Drexler and P. Elbau]%
		[]%
		{}%
		{\footnotesize\sc A Quantitative Model for Optical Coherence Tomography}%
		{\footnotesize\thepage}
	\setfoot{}{}{}
}
\usepackage{dsfont}
\usepackage{bm}
\usepackage{siunitx}

\newcommand{\R}{\mathds{R}}
\newcommand{\C}{\mathds{C}}

\let\RE\Re
\let\Re=\undefined
\DeclareMathOperator{\Re}{\RE e}

\DeclareMathOperator{\sinc}{si}
\let\IM\Im
\let\Im=\undefined
\DeclareMathOperator{\Im}{\IM m}

\DeclareMathOperator{\argmax}{argmax}
\DeclareMathOperator{\argmin}{argmin}
\DeclareMathOperator{\sign}{sign}
\DeclareMathOperator{\erfii}{erfi}

\usepackage{chngcntr}
\counterwithout{equation}{section}

\usepackage{standalone}
\usepackage{caption}
\usepackage{subcaption}

\begin{document}

\maketitle
\thispagestyle{empty}
\begin{center}
\hspace*{5em}
\parbox[t]{12em}{\footnotesize
\hspace*{-1ex}$^1$University of Vienna\\
Oskar-Morgenstern-Platz 1\\
A-1090 Vienna, Austria}
\hfil
\parbox[t]{17em}{\footnotesize
\hspace*{-1ex}$^2$Medical University of Vienna\\
Waehringer Guertel 18-20\\
A-1090 Vienna, Austria}
\end{center}

\begin{abstract}
Optical coherence tomography (OCT) is a widely used imaging technique in the micrometer regime, which gained accelerating interest in medical imaging 
in the last twenty years. In up-to-date OCT literature \cite{brezinski_optical_2006,drexler_optical_2008} certain simplifying assumptions are made for the reconstructions, but for many applications a more realistic description of the OCT imaging process is of interest. 
In mathematical models, for example, the incident angle of light onto the sample is usually neglected or 
a plane wave description for the light-sample interaction in OCT is used, which ignores almost completely the occurring effects within an OCT measurement process. 
In this article, we make a first step to a quantitative model by considering the measured intensity as a combination of back-scattered Gaussian beams affected by the system. In contrast to the standard plane wave simplification, the presented model includes system relevant parameters such as the position of the focus and the spot size of the incident laser beam, which allow a precise prediction of the OCT data and therefore ultimately serves as a forward model. The accuracy of the proposed model---after calibration of all necessary system parameters---is illustrated by simulations and validated by a comparison with experimental data obtained from a \SI{1300}{\nano\metre} swept-source OCT system.
\end{abstract}

\section{Introduction}
Optical coherence tomography (OCT) has proved to be a non-invasive, high-precision imaging technique with micrometer resolution. It emerged around 1990 for in-vivo imaging of the human eye \cite{fercher_eye-length_1988, huang_optical_1991} and gained increasing interest ever since. Nowadays, extensions like angiography \cite{khan_major_2017}, polarization sensitive OCT \cite{baumann_polarization_2017} and optical coherence elastography \cite{zaitsev_strain_2021} unlocked a wide range of possible applications; for example, blood vessel analysis \cite{liu_optical_2019} and cancer margin detection \cite{kennedy_diagnostic_2020}, while OCT endoscopes \cite{albrecht_vivo_2020} are clearing the way for high resolution imaging of internal organs and their pathologies. Multi-modal imaging techniques \cite{schie_morpho-molecular_2021} often use OCT as morphological guidance.

While many theoretical OCT articles assume a sample geometry that is perfectly perpendicular to the OCT beam \cite{AndThrYurTycJorFro04, FenWanEld03, RalMarCarBop06}, commonly used OCT systems are designed for rough sample surfaces and arbitrary sample inclinations, which yield much less power at the detector. Normal incidence not only oversaturates the detector easily, especially for samples with a high refractive index and a very directed scattering profile, but can also lead to interference between the sample and optical parts inside the setup, e.g. the scan lens. To prevent imaging artifacts, normal incident is therefore usually avoided in OCT. 

In addition, most works are based on a plane wave ansatz for describing the sample and reference fields \cite{FerDreHitLas03, izatt2008theory, Kal17, TomWan05}. While this approximation is valid for the immediate focus region, it is most often not true for the whole field of view of the setup or even the whole sample area. Common effects like a focus dependent intensity profile inside the sample cannot be described with a plane wave ansatz. While workarounds like multiplying the spectral resolution with a sensitivity factor have been proposed \cite{DreFuj15}, in this work a Gaussian beam, similar to \cite{BreMunKrueKie19}, is used as incoming wave for a more precise description of the light beams.

To make a quantitative reconstruction of the optical parameters of the sample possible, we make the simplifying assumption that the sample is not absorbing and can at least locally be described as a layered medium (with layers not necessarily perfectly perpendicular to the incident light), which is a classical assumption in this field. 
This simplification allows us to analytically calculate the scattered light from the sample, which is then collected by a scan lens and combined with the reference light to produce the interference pattern. We roughly model the effect of the scan lens on the scattered light by discarding plane wave components moving in wrong directions. Together with the layers and the focusing effects introduced by Gaussian beams we derived simulations which have been in much better accordance with the experimental data obtained by a \SI{1300}{\nano\metre} swept-source OCT system compared to a simple plane wave approach.

The paper is structured as follows: In Section~\ref{sec:experiments} we describe the OCT imaging system that was built for this work. The individual experiments performed for investigating the different effects of the system on the data are also introduced. In Section~\ref{sec:math_model}, we present the general problem and the governing equations. We give the forms of the sample and reference fields using the near- and far-field representation of back-scattered Gaussian waves. The section ends with the derivation of the formula for the measurement data. In Section~\ref{sec:calibration} we present in two experiments the dependence of the data on the incident angle and the beam focusing and show how to use them as calibration tools to determine otherwise unknown parameter such as the beam radius in the focus. The comparison between experimental and simulated data is presented in the last section. There, we see that our model nicely predicts the behavior of the experimental data with respect to different orientations and positions of the sample relative to the focus.

\section{Experiments}\label{sec:experiments}

Since there are many different variants of OCT systems around, we briefly describe the system we use for generating the data and the individual experiments. 
\subsection{OCT Setup and Post-Processing}\label{subsec:setup}

\begin{figure}[hbt!]
    \centering
    \includegraphics[width=0.7\textwidth]{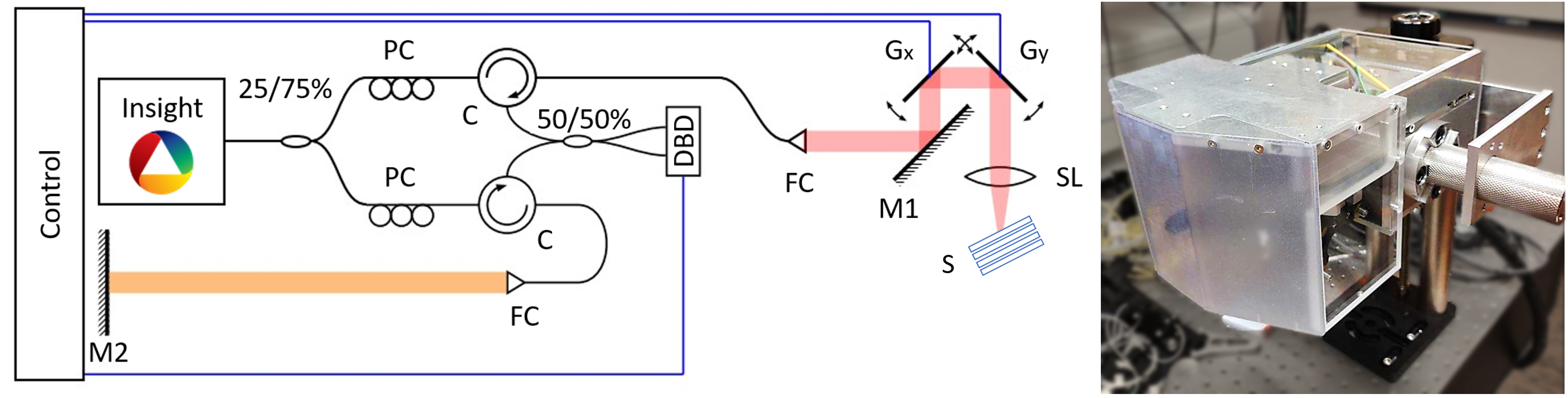}
    \caption{OCT setup: Insight: \SI{1300}{\nano\metre} swept-source; 25/75~\% fiber coupler; PC: polarization control; C: circulator; 50/50~\% fiber coupler for light recombination; FC: fiber collimator; M1, M2: mirrors; Gx, Gy scanning galvanometers; SL: scan lens; S: sample; DBD: dual-balance detector. 
    } \label{fig:octsystem}
\end{figure}

All measurements where performed with a custom-built fiber-based OCT system with a central frequency of about \SI{1300}{\nano\metre} and \SI{30}{\nano\metre} bandwidth, schematically shown in Figure~\ref{fig:octsystem}. The core of the setup is the akinetic swept-source from Insight Photonic Solutions, USA, which emits about \SI{60}{\milli\watt} at a repetition rate of up to \SI{500}{\kilo\hertz}. This swept-source shows a flat power profile over its whole bandwidth and a high phase stability, making it well-suited for any type of signal analysis. A fiber optics coupler guides 75~\% of the laser light into the sample arm and 25~\% into the reference arm, where a \SI{4}{\milli\metre} fiber collimator releases it onto a short free-space path, with a moveable mirror at the end, that reflects the light back into the collimator. The custom-built sample arm features a rotatable imaging probe with conjugated scanning and a LSM54-1310 scan lens from Thorlabs, USA, for a flat imaging plane. Circulators are used to guide the reference and sample arm signal to a 50/50~\% fiber coupler, where the laser light recombines. A dual-balance-detector (BPD-1, Insight Photonic Solutions, USA), short DBD, records the cross-correlation term and an ATS9360 data acquisition card from Alazar Technologies, Canada is used to digitalize it.

The Insight source supplies a trigger signal, which is used by a field programmable gate array to coordinate galvanometer movement and triggers the data acquisition card. Since the swept-source has a 100~\% duty cycle, which the data acquisition card can not keep up with, every second sweep is neglected, to preserve the whole spectrum for imaging. An attenuation wheel in the reference arm free-space beam path is used to control the interference power detected by the DBD and thereby the intensity of the recorded OCT signal. It is used to ensure a high signal, without oversaturation of the detector. The system achieves an axial resolution of \SI{31}{\micro\metre} and a lateral resolution of \SI{24}{\micro\metre} in air as well as an SNR of \SI{105}{\decibel}. The OCT control software was written in Labview and data processing was performed in MATLAB.

The recorded spectrum with 700 datapoints can immediately be Fourier transformed into image space, since the wavelength sweep emitted by the laser is already spaced equally in terms of wavenumber. The small bandwidth makes dispersion compensation unnecessary. Background removal is performed via subtraction of the average spectrum of a volume. Zeropadding ensures the small axial pixelsize of \SI{13.7}{\micro\metre} in air and the lateral pixelsize is \SI{9.8}{\micro\metre} in air.
The DBD records only the difference between both input signals, thereby removing common-mode noise and centering the signal around zero. Through the digitization process the signal is shifted in height so it can be stored as unsigned integer. This shift is removed again during post-processing through the background subtraction.

\subsection{Power vs. Angle}\label{subsubsec:data_angle}\label{subsec:data_prep}
To investigate the influence of the incident angle between sample surface and OCT beam, one of the fibers entering the DBD was connected to an optical power meter (PM100C with S122C, Thorlabs, USA) instead. A mirror was fixed onto a goniometer stage with a Vernier scale (GOH-65A100RUU, OptoSigma, USA) with a kinematic mount. First, the goniometer was aligned to ensure that the OCT beam goes through its Pivot point. Second, the kinematic mount and a vertical stage were adjusted to put the mirror in focus and ensure normal incident of the laser beam on the mirror, using the power meter as guidance. Then the mirror was tilted in 5 arc minute steps back and forth between \SI{-1}{\degree} and \SI{1}{\degree} and the power was recorded until each angular position was measured $6$ times.  

\subsection{Power vs. Focus}\label{subsubsec:data_focus}
For quantification of the Gaussian behavior of the focus a motorized stage (T-LSM050A, Zaber Technologies, Canada) was used to transport once a mirror and once a microscopy coverglass through the focus of the OCT system, with a roughly fixed tilt of about \SI{2.75}{\degree}. The coverglass (631-0124, VWR International, USA) has a refractive index of $1.5088$ for \SI{1300}{\nano\metre}, which needs to be taken into account during data analysis. At each position of the stage, we use $11$ steps for the mirror and $7$ for the coverglass, a 3D OCT volume was recorded. These 3D volumes were post-processed according to Section~\ref{subsec:setup} and used to determine the exact incident angle and the distance of the center of the sample surface to the position where sample and reference arm would have the same length, called zero delay.

\section{Mathematical Model}
\label{sec:math_model}
\begin{figure}[hbt!]
\centering
\begin{tikzpicture}[x=1.17cm,y=1.17cm,font=\scriptsize]
\node  at (6,-2.17) {\includegraphics[width=14cm]{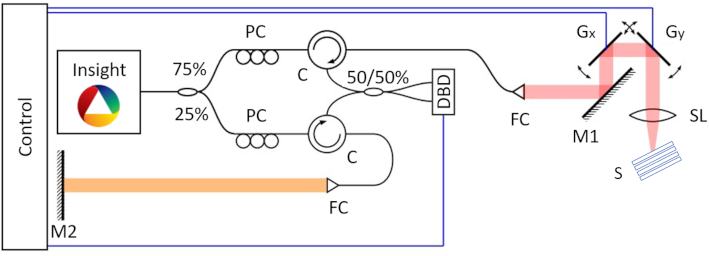}}; 
\filldraw[white,opacity=0.4](0,0) rectangle (12,-4.34);

\filldraw[green,thick,fill opacity=0.1](0.83,-2.42) rectangle +(4.59,-1.68) node[right,anchor=south east,fill opacity=1] {$\bm E_{\mathrm R}^{(0)}$, $\bm E_{\mathrm R}$, $\bm E_{\mathrm R}^{(1)}$, \autoref{se:referenceField}};

\filldraw[red,thick,fill opacity=0.1](0.83,-0.43) rectangle +(2.08,-1.95);
\draw[red](0.8,-0.38) node[anchor=north west] {$\bm E^{(0)}$, \autoref{se:Gaussian}};

\filldraw[orange,thick,fill opacity=0.1](10.15,-2.19) rectangle +(1.85,-1.8);
\draw[orange](10.15,-2.10-1.8) node[anchor=south west] {\parbox{1.8cm}{$\bm E_{\mathrm S}^{(0)}$, $\bm E_{\mathrm S}$\\\autoref{se:sampleField}}};

\filldraw[blue,thick,fill opacity=0.1](5.88,-0.27) rectangle +(1.95,-1.81);
\draw[blue](5.88,-0.27) node[anchor=north west] {\parbox{1.52cm}{$\bm E_{\mathrm S}^{(1)}+\bm E_{\mathrm R}^{(1)}$\\\autoref{se:dbd}}};

\definecolor{citrine}{rgb}{0.89, 0.82, 0.04}
\filldraw[citrine,thick,fill opacity=0.1](10.4,-1.20) rectangle +(2.95,-0.91);
\draw[citrine](10.45+2.95,-1.20) node[anchor=north east] {$\bm E_{\mathrm S}^{(1)}$, \autoref{se:scanlens}};
\end{tikzpicture}
\caption{Modeling of the separate parts of the OCT experiment: We start by describing the light beam produced by the laser (the red box) in Section~\ref{se:Gaussian}, we give a representation for the beam in the sample arm which is backscattered from the sample (the orange box) in Section~\ref{se:sampleField} and is then coupled back into the fiber system via the scan lens (the yellow box) in Section~\ref{se:scanlens}. This is afterwards recombined with the beam from the reference arm (the green box) in Section~\ref{se:referenceField} and detected by the dual balance detector (the blue box) in Section \ref{se:dbd}.}
\label{fig:model}
\end{figure}
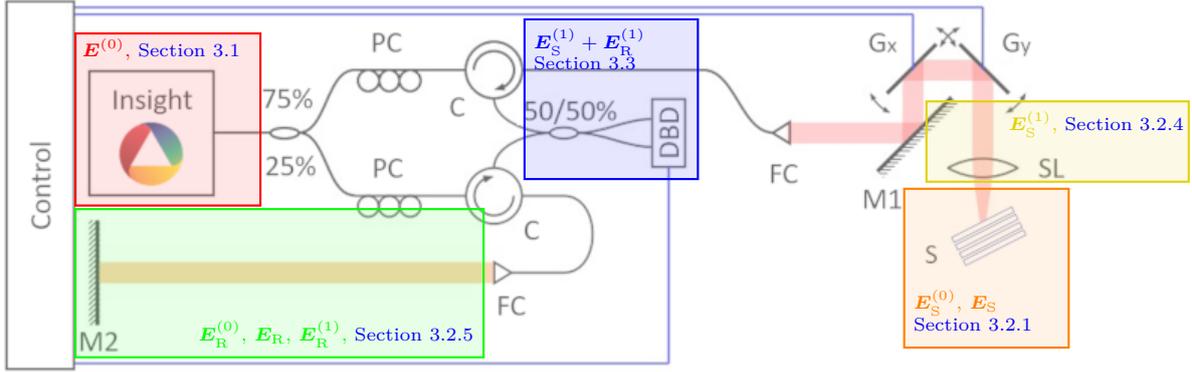

Considering the workflow of the used OCT system, described in the previous section, we model the parts shown in Figure~\ref{fig:model} separately.

Firstly, in Section~\ref{se:Gaussian}, we model the produced laser illumination. The laser light is split into two beams, one is sent to the sample and the other to the mirror in the reference arm. We model their scattering process in the Sections~\ref{se:sampleField} and \ref{se:referenceField} respectively.  

The reflected light in the sample arm is (partially) collected by a scan lens and coupled into a fiber. This aspect is the topic of Section~\ref{se:scanlens}.         

After recombination of the scattered light beams, we model the detection via a dual-balancing detector of this superposition in Section~\ref{se:dbd}. This in particular is discussed for the measurement related to two experiments explained in Section~\ref{sec:experiments}, which in the end are obligatory for the calibration of necessary parameters in the forward simulations.          

\subsection{Gaussian Beam illumination}\label{se:Gaussian}
The shape of the light produced inside an optical resonator (we ignore at this point the finite size of the resonator and the boundary conditions) of a laser can according to \cite{Sve10}, for example, be well described by a Gaussian beam. 

We consider a Gaussian beam $\mathbf E\colon\R^3 \to \C^3$ as a monochromatic solution of the electromagnetic wave equation in vacuum which reduces it to Helmholtz equation (usually it is considered as solution of the paraxial approximation of the wave equation which is not necessary here): 
\begin{equation} \label{eq:Helmholtz_system}
\begin{aligned}
\Delta \mathbf E(\mathbf x)+k_0^2\mathbf E(\mathbf x) &= 0, & \mathbf x\in\R^3,\\
\langle\nabla,\mathbf E\rangle(\mathbf x) &= 0, & \mathbf x\in\R^3. 
\end{aligned}
\end{equation}  
It is characterized by its form
\begin{equation}
\label{eq:focal_plane}
\mathbf E(x_1,x_2,r_0) =  f(x_1,x_2)\mathbf p
\end{equation} 
in the focal plane $\{\mathbf x\in\R^3\ |\ x_3 = r_0\}$ for a function $f:\R^2\to \C$ such that its 2D Fourier transform is compactly supported in $D_{k_0}(0)$ (the open ball with center $0$ and radius $k_0$) and a polarization vector $\mathbf p\in\R^2\times\{0\}.$ 

\begin{theorem}
\label{thm:thm1}
Let $f:\R^2\to\C$ be a function such that its two-dimensional Fourier transform $\check f$ is compactly supported in $D_{k_0}(0)$ and let $\mathbf p\in\R^2\times\{0\}$. Then for every $\mathbf x\in\R^3$ a solution of the Helmholtz problem \eqref{eq:Helmholtz_system} is given by 
\begin{multline}
\label{eq:incident_gaussian}
\mathbf E(\mathbf x)=\frac{1}{4\pi^2}\int_{\R^2} \check{\bm g}(k_1,k_2) e^{-i (k_1 x_1 + k_2 x_2)}e^{ -i \sqrt{k_0^2-(k_1^{2}+k_2^{2})}(r_0-x_3)} d (k_1,k_2)\\
-\frac{1}{4\pi^2}\int_{\R^2} \check{\bm g}(k_1,k_2) e^{-i (k_1 x_1 + k_2 x_2)}e^{ i \sqrt{k_0^2-(k_1^{2}+k_2^{2})}(r_0-x_3)} d (k_1, k_2),
\end{multline}
with 
\begin{equation}
\label{eq:initial_dist}
\check{\bm g}(k_1,k_2) = \frac{1}{2}\check f(k_1,k_2) \begin{pmatrix} p_1\\ p_2\\ \frac{p_1 k_1+ p_2 k_2}{\sqrt{k_0^2-(k_1^{2}+k_2^{2})}} \end{pmatrix}.
\end{equation}
\end{theorem}

Such a wave describes well the light inside the optical resonator of the laser. Then through one partly transparent mirror of the resonator, we then obtain only the light moving in the negative $x_3-$direction of the form
\begin{equation}
\label{eq:incident_gaussian_sample}
\mathbf E^{(0)}(\mathbf x)=\frac{1}{4\pi^2} \int_{\R^2} \check {\bm g}(k_1,k_2) e^{-i (k_1 x_1 + k_2 x_2)} e^{-i\sqrt{k_0^2-(k_1^{2}+k_2^{2})}(r_0-x_3)}d (k_1,k_2).
\end{equation}
Hereby, a reasonable model for the shape of the function $f$ is one which resembles a Gaussian function.

This laser light is transported within single-mode fibers through the OCT system, conserving the shape of the Gaussian beam throughout the system.

\subsection{Backscattered Gaussian Fields}

The laser light is split into two waves, $\mathbf E_S^{(0)}$ for the sample and $i\mathbf E_R^{(0)}$ for the reference arm, as in \eqref{eq:incident_gaussian_sample} respectively, by a beam splitter and both remain in the form of a Gaussian beam, with possibly different beam parameters, for example due an optical attenuation wheel inside the reference arm, which causes a difference in light intensities between the beams.

\subsubsection{The Sample Field}\label{se:sampleField}

The beam $\mathbf E_S^{(0)}$ is now directed onto the sample and we say it is of form \eqref{eq:incident_gaussian_sample} with $f=f_S.$ Then, if the beam is sufficiently focused, meaning that the values of $|\mathbf E^{(0)}_S|$ can be neglected outside a small region, we only need to consider for the scattering process the shape of the sample inside this region. In this subregion, we denote it by $\Omega$, we assume, using the tangent plane approximation, that it can be described by a layered structure. These layers are not necessarily perpendicular to incident beam, but for simplification assumed to parallel to each other. This is modeled by $\Omega$ being a finite union of layers: 
\[
\Omega=\bigcup_{j=1}^L\Omega_j,\quad \Omega_j=\{\mathbf x\in\Omega\ |\ a_j\leq\langle \mathbf x,\bm\nu_\Omega\rangle<a_{j+1}\}, \quad (a_j)^{L}_{j=1}\subset\R, 
\] 
for some unit normal vector $\bm\nu_\Omega.$ Each of these shall be characterized by a constant refractive index $n_j\in[1,\infty).$

Under these conditions, we model the backscattered field $\mathbf E_S$ as solution of Helmholtz equation
\begin{equation}
\label{eq:scat_prob}
\Delta (\mathbf E_S + \mathbf E_S^{(0)})(\mathbf x) + k_0^2 n^2(\mathbf x) (\mathbf E_S + \mathbf E_S^{(0)})(\mathbf x) = 0, \quad \mathbf x\in\R^3
\end{equation}
where $n(\mathbf x) = \sum_{j=1}^L n_j \chi_{\Omega_j}(\mathbf x) + \chi_{\R^3\setminus\Omega}(\mathbf x)$ and appropriate radiation conditions are assumed.

The incident field \eqref{eq:incident_gaussian_sample} is represented as a superposition of plane waves having different wave vectors. Because of the linearity of the equation it is sufficient to solve the problem for every plane wave. The result for these backscattered fields for such a sample is well known in this plane wave case, see \cite{Jac98,ElbMinVes20}. For the simplest case $L=1,$ we consider an (arbitrary) plane wave as incident illumination from the top,  
\[
\mathbf E_{S}^{(0),\text{pl}}(\mathbf x) = \bm \alpha(k_1,k_2) e^{-i \langle\mathbf k, \mathbf x\rangle}, \quad \mathbf x\in\R^3,
\]
with amplitude function ${\bm \alpha}:\R^2\to\C^3$ 
and propagation vector 
\begin{equation}
\label{eq:prop_vec}
\mathbf k=\begin{pmatrix}k_1\\k_2\\-\sqrt{k_0^2-k_1^2-k_2^2}\end{pmatrix}, \ |\mathbf k| = k_0,
\end{equation}
which we consider implicitly as a function of $k_1$ and $k_2.$ We obtain the reflected electric field 
\begin{equation}
\label{eq:solution}
\mathbf E_S^{\text{pl}}( \mathbf x) =\bm \beta(k_{1},k_{2}) \bm \alpha(k_1,k_2) e^{-i\langle (\mathbf k-\mathbf k_r),\mathbf x_\Omega\rangle}e^{-i \langle\mathbf k_r, \mathbf x\rangle}, 
\end{equation}
where $\mathbf x_\Omega$ denotes an arbitrary point of the top boundary (that is $\langle \mathbf x_\Omega,\bm\nu_\Omega\rangle = a_1 $) of the object, 
\begin{equation}
\label{eq:refvec}
\Phi:\R^3\to \R^3,\quad \mathbf k_r= \Phi(\mathbf k)=\mathbf k -2 \langle \mathbf k,\bm\nu_\Omega \rangle \bm\nu_\Omega
\end{equation} 
the wave vector and $\bm\beta_S$ the sum of the reflection coefficients $\beta_{0,\parallel},\beta_{0,\perp}$ of the differently polarized parts
\begin{equation}
\label{eq:full_reflection_coefficient}
\bm \beta_S(k_{1},k_{2}) = \lambda_1\beta_{0,\parallel} \mathbf p_{\parallel}(k_1,k_2) 
+ \lambda_2 \beta_{0,\perp} \mathbf p_{\perp}(k_1,k_2).
\end{equation}
Here, we have decomposed $\bm \alpha$ into its transverse electric and magnetic polarizations, with coefficients $\lambda_1$ and $\lambda_2,$ respectively. Further, we use Snell's law for the determination of the transmission angle $\theta_t.$

Summarizing the scattered (plane) waves for all $(k_1,\,k_2)$ and \[{\bm \alpha}(k_1,k_2) = \check{\bm g}_S(k_{1},k_{2}) e^{-i \sqrt{k_0^2 - k_{1}^2 -k_{2}^2}r_0}\] then finally results in
\begin{align}
\mathbf E_S(\mathbf x) &= \frac{1}{4\pi^2} \int_{\R^2}\mathbf E_{S}^{\text{pl}}(\mathbf x)d (k_1,k_2) \nonumber\\
&= \frac{1}{4\pi^2} \int_{\R^2} \bm\beta_S(k_1,k_2) \check f_S(k_1,k_2) e^{-i\sqrt{k_0^2-k_1^2-k_2^2}r_0} e^{-i \langle(\mathbf k-\mathbf k_r), \mathbf x_\Omega\rangle} e^{-i \langle\mathbf k_r, \mathbf x\rangle} d (k_1,k_2),\label{eq:samplefield_near}
\end{align}
with $\bm\beta$ given by \eqref{eq:full_reflection_coefficient}.

\subsubsection{Far Field Method}

Since the distance between the scan lens and the sample (which is roughly \SI{6}{\centi\metre}) is much greater than the size of of the sample itself (which is only a few millimeters), we can be tempted to simplify the integrand by using the far-field approximation.    
 
Mathematically, this means, that we are approximating \eqref{eq:samplefield_near} by its behavior at some point $r \mathbf s, \ \mathbf s\in\mathbb S^2,$ as $r\to\infty:$ 
\[
\mathbf E_S(r\mathbf s) = \mathbf E_{S,\infty}(r\mathbf s) + o(1/r) 
\]

To compute the dominating term $ \mathbf E_{S,\infty}$, we apply the method of stationary phase, see Lemma \ref{lem:stationary_phase}, which is based on the approximation of the phase function $k_0\Psi,$ with 
\[
\Psi(k_1,k_2)= \langle \tfrac{\mathbf k_r}{k_0 }, \mathbf s\rangle 
             = \langle \tfrac{\mathbf k}{k_0 }, \mathbf s\rangle - 2 \langle \tfrac{\mathbf k}{k_0 }, \nu_\Omega\rangle \langle \nu_\Omega, \mathbf s\rangle,  
\]
by its Taylor series around its critical points. 

\begin{theorem}
\label{thm:first_order}
Let $\mathbf E_S$ be a vector field given by \eqref{eq:samplefield_near}. Then, its far field approximation takes the form
\begin{equation}
\label{eq:farfield}
\mathbf E_{S,\infty}(r\mathbf s) =  \frac{ -i k_0 \left| c_3 \right| }{2\pi r} \bm\beta_S(k_1,k_2)\check f_S(k_1,k_2)  e^{-i\sqrt{k_0^2-k_1^2-k_2^2}r_0} e^{-i \langle \mathbf k - \mathbf k_r,\mathbf x_\Omega\rangle}e^{i k_0 \sign(c_3)r },
\end{equation}
where for $c_3 = s_3-2 \langle \nu_\Omega, \mathbf s \rangle \nu_{\Omega,3},$ the vector components $k_1$ and $k_2$ are given by 
\begin{equation}
\label{eq:dir}
\begin{pmatrix}k_1 \\ k_2 \end{pmatrix} = -k_0\sign(c_3) \begin{pmatrix} s_1 - 2\langle \nu_\Omega, \mathbf s \rangle \nu_{\Omega,1}\\ s_2 - 2\langle \nu_\Omega, \mathbf s \rangle \nu_{\Omega,2} \end{pmatrix},
\end{equation}
and for $k_1, k_2$ the reflected vector $\mathbf k_r=\Phi(\mathbf k)$ is given by \eqref{eq:refvec}.   
\end{theorem}

\subsubsection{Comparing the Near- and the Far-Fields}
\label{subsec:near_far}

Comparing the representations \eqref{eq:samplefield_near} and \eqref{eq:farfield} of the scattered electric field simulations make it obvious that there is a difference in the visible effects provided by these methods. In this work are concerned with the influence of the focus in the scattered field. 

\begin{figure}[hbt!]
\centering
\includegraphics[width=0.65\textwidth]{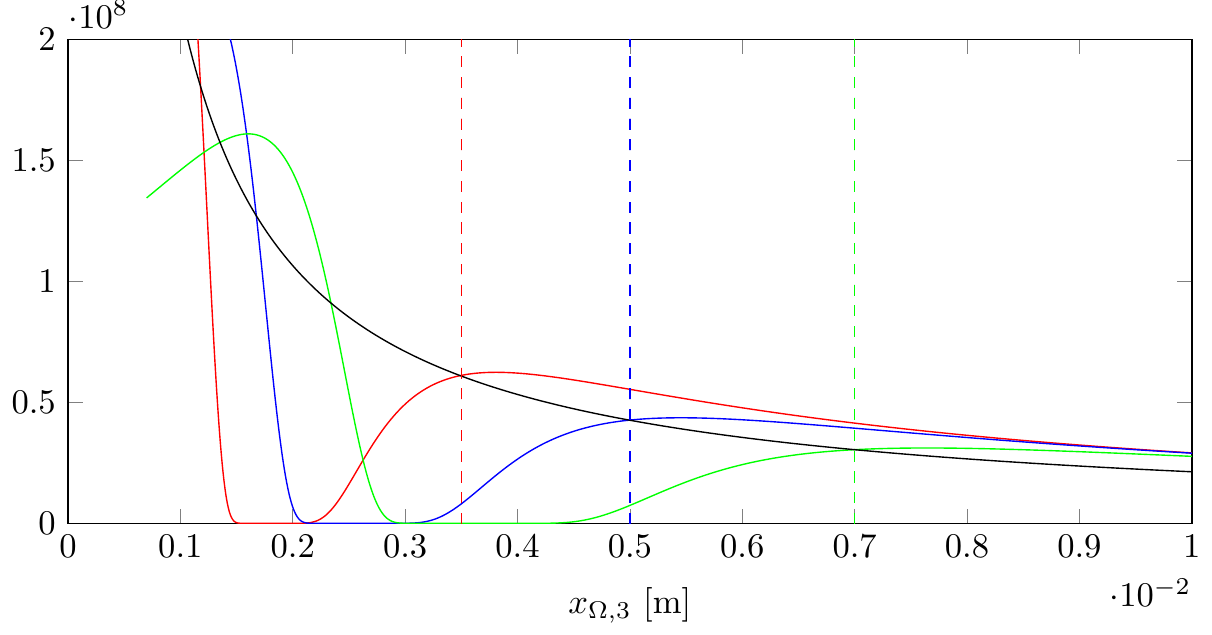}
\caption{The near-field for different positions of the focus (red, blue, green) vs. the far-field (black) regime for different positions of the surface. At the dotted lines, indicating where the surface and focus position coincide, the intensities of near- and far-field regime are equal. } \label{fig:near_far}
\end{figure}

In contrast to the far-field representation, the scattered field in the near field regime is heavily depending on the distance between the positions of the layer and of the focus. We neglect the vectorial quantities in \eqref{eq:samplefield_near} for a moment and allow for an amplitude function 
\begin{equation}\label{eq:fexp}
\check f(k_1,k_2) \sim e^{-a(k_1^2+k_2^2)},
\end{equation} 
where the parameter $a>0$ is such that the error $|\check f-\check f\chi_{D_{\rho_0}(0)}|$, is negligible. Hereby, $D_{\rho_0}(0)$ is a disk with small radius $\rho_0$ and center $0.$ Then, for a single surface (medium) we obtain the scattered field 
\[
\mathbf E(\mathbf x) = \frac{1}{4\pi^2} \int_{\R^2} \beta_{0}(k_1,k_2) e^{-a(k_1^2+k_2^2)} e^{-i\sqrt{k_0^2-k_1^2-k_2^2}r_0} e^{-i \langle(\mathbf k-\mathbf k_r), \mathbf x_\Omega\rangle} e^{-i \langle\mathbf k_r, \mathbf x\rangle} d (k_1,k_2).
\]
We assume that on the small disk, the reflection coefficient $\beta_0$ is approximately constant and due to small deviations of $k_1,k_2$ from zero we may approximate the root in the exponents
\begin{equation}
\label{eq:small_deviation}
\sqrt{k_0^2-k_1^2-k_2^2} \simeq k_0 - \frac{k_1^2+k_2^2}{2k_0}. 
\end{equation}       
Further, for the sake of simplification, we restrict $\bm \nu_\Omega\in\R^2\times\{0\}, \nu_{\Omega,2}=0,$ fix the positions of the focus $ r_0$ and the object $\mathbf x_\Omega = x_{\Omega,3},\ x_{\Omega,3} <0$ below the origin and evaluate at $\mathbf x=\mathbf 0.$
This then finally gives for the scattered field
\begin{equation}
\label{eq:simple}
\mathbf E(\mathbf 0) = \frac{\beta_0}{4\pi^2} e^{-ik_0 \psi_0} \int_{\R} e^{-\psi_2 k_1^2} e^{-i k_1  \psi_1} d k_1  \int_\R e^{-\psi_2 k_2^2 } d k_2,
\end{equation}
where we defined the phase elements $\psi_0,\psi_1,\psi_2$ by
\begin{equation}
\label{eq:parameters}
\psi_2(k_0) = a+\frac{i}{k_0}\left( \nu_{\Omega,3}^2 x_{\Omega,3} -\frac{r_0}{2}\right),\quad \psi_1 = 2\nu_{\Omega,1}\nu_{\Omega,3}x_{\Omega,3},\quad \psi_0=r_0-2\nu_{\Omega,3}^2x_{\Omega,3}. 
\end{equation}
Since $\Re(\psi_2) = a >0,$ we evaluate both integrals in \eqref{eq:simple} and arrive at  
\[
\mathbf E(\mathbf 0) = \frac{\beta_0}{4 \pi \psi_2(k_0) } e^{-\frac{\psi_1^2}{4\psi_2(k_0)}} e^{-ik_0 \psi_0}. 
\]
After complex conjugation in the exponent and taking the absolute value of the field, we find that 
\begin{equation}
\label{eq:abs_near_field}
|\mathbf E(\mathbf 0)| = \frac{|\beta_0| k_0 }{4\pi \sqrt{k_0^2 a^2 + d^2}} e^{-\frac{k_0^2 \psi_1^2 a }{4(k_0^2 a^2 +  d ^2) }},
\end{equation}
with distance $d = \nu_{\Omega,3}^2 x_{\Omega,3} -\frac{r_0}{2}.$ Considering now \eqref{eq:abs_near_field} for different positions $x_{\Omega,3}$ of the sample, and therefore for varying $d,$ corresponds to different evaluation points $\mathbf x_\infty=r\mathbf e_3,$ with $r=|x_{\Omega,3}|$ in the far-field regime. Taking the absolute value of \eqref{eq:farfield} 
\[
|\mathbf E_\infty(r \mathbf e_3)| \simeq \frac{k_0|\beta_0| |c_3|}{2\pi r} e^{-a(k_1^2+k_2^2)}
\]         
we observe that opposed to the near-field representation, the far-field regime is independent of the focus position. Figure \ref{fig:near_far} provides a comparison between both for different positions of the focus and the surface.

Since the far-field approximation does not show the dependence of the electric field on the focus, we stick with the more complicated near-field representation of the scattered light $\mathbf E_S$ in \eqref{eq:samplefield_near}.

\subsubsection{The Scan Lens}
\label{se:scanlens}
 
The backreflected light $\mathbf E_S$ then passes trough the scan lens and is collected by a fiber collimator. Thereby, we loose all plane wave components whose propagation directions are outside a certain angular range of the collimator. We model this by reducing the area of integration in \eqref{eq:samplefield_near} to a set $\mathcal B$ of those scattered wave directions $\mathbf k_r$ which have an angle to the measurement direction $\mathbf e_3$ less than a certain angle of acceptance $\theta_{max},$ that is
\begin{equation}
\label{eq:int_domain}
\mathcal B = \left\{ (k_1,k_2)\in\R^2 \ |\ \arccos\left(\frac{\langle \mathbf k_r,\mathbf e_3\rangle}{k_0} \right) \leq \theta_{max} \right\}, 
\end{equation}
which finally gives a (scattered) sample field 
\begin{equation}
\label{eq:samplefield}
\mathbf E_S^{(1)}(\mathbf x_\mathcal D) = \frac{1}{4\pi^2} \int_{\mathcal B} \bm\beta_S(k_1,k_2) \check f_S(k_1,k_2) e^{-i\sqrt{k_0^2-k_1^2-k_2^2}r_0} e^{-i \langle(\mathbf k-\mathbf k_r), \mathbf x_\Omega\rangle} e^{-i \langle\mathbf k_r, \mathbf x_\mathcal D\rangle} d (k_1,k_2).
\end{equation}
at the scan lens position $\mathbf x_\mathcal D = r_\mathcal D \mathbf e_3$ above the sample.

\subsubsection{The Reference Field}\label{se:referenceField}

Similarly, we model the reference field as solution to the scattering problem \eqref{eq:scat_prob} with Gaussian incident illumination $\mathbf E_R^{(0)}$ and a medium with constant (infinitely) large refractive index. Following the same line that led to \eqref{eq:samplefield_near}, we get with $f=f_R,$ a field of the form
\[
\mathbf E_R(\mathbf x_\mathcal D) = \frac{1}{4\pi^2} \int_{D_{k_0}(0)} \bm \beta_R(k_1,k_2)\check f_R(k_1,k_2) e^{-i\sqrt{k_0^2-k_1^2-k_2^2}r_0} e^{-i \langle(\mathbf k-\mathbf k_r), \mathbf x_M\rangle} e^{-i \langle\mathbf k_r, \mathbf x_\mathcal D\rangle} d (k_1,k_2)
\] 
Following the experimental setup the mirror in the reference arm is perpendicular to the incident light, so that the unit normal vector $\bm \nu_M = \mathbf e_3,$ and positioned in the focus of the $\mathbf E_R^{(0)}$, such that, following Section~\ref{subsec:near_far}, the far-field approximation for reference field $\mathbf E_R$ is valid. We thus have a reference field $\mathbf E_R^{(1)},$ given by
\begin{equation}
\label{eq:reference_field}
\mathbf E_R^{(1)}(r_\mathcal D\mathbf e_3) = \frac{ -i k_0 }{2\pi r_\mathcal D} \bm \beta_R(0,0) \check f_R(0,0)  e^{-i k_0 (r_0+r_\mathcal D-2x_{M,3})}. 
\end{equation}

\subsection{OCT Measurements}
\label{se:dbd}

With the identities of the incident and the backscattered light in hand, we now proceed with the modeling of the detection process inside an OCT system. 

In order to suppress common-mode noise and enhance the signal-to-noise ratio, a dual-balance-detector is used to record the OCT signal. After recombination of sample and reference arm light, the laser signal is split 50/50$\%$ into two different fibers $\mathbf F_1,\mathbf F_2$, each entering one of the DBDs optical inputs. The DBD then subtracts one input from the other, thereby removing everything but the cross-correlation term of the interference.

Thus, assuming that the sample and the reference fields $\mathbf E_S^{(1)}$ and $i \mathbf E_R^{(1)}$ are passing through a perfect splitter, we obtain the forms for the fields in the fibers as  
\begin{equation*}
\mathbf F^1 = \frac{1}{\sqrt{2}}\left( \mathbf E_S^{(1)} -\mathbf E_R^{(1)}\right), \quad \mathbf F^2 = \frac{1}{\sqrt{2}}\left( i\mathbf E_S^{(1)} + i \mathbf E_R^{(1)}\right).
 \end{equation*} 
We assume, ignoring the travel paths inside the fibers, that these fields are detected at the position $\mathbf x_\mathcal D$ of the scan lens. These measurements are performed for different wavenumbers $k_0$ in a scan range $[k_{min},\, k_{max}].$ We therefore indicate explicitly the dependence on $k_0$ in the measurements: 
\begin{equation}
\label{eq:dbd_data2}
\mathcal M(k_0) =  \frac{1}{2} \left(|\mathbf F^1 (x_\mathcal D)|^2 - |\mathbf F^2 (x_\mathcal D)|^2\right) = - \Re\left\langle\mathbf E_S^{(1)},\overline{\mathbf E_R^{(1)}}\right\rangle, \quad k_0\in[k_{min},\, k_{max}].
\end{equation}
With the identities \eqref{eq:samplefield} and \eqref{eq:reference_field} for $\mathbf E_S^{(1)}$ and $\mathbf E_R^{(1)},$ we obtain  
\begin{equation}
\label{eq:cro}
\mathcal M(k_0) = \frac{k_0 }{8 r_\mathcal D \pi^3 } \int_{\mathcal B} \Re\left(-i \left\langle \bm \beta_S(k_1,k_2)\check f_S(k_1,k_2), \bm \beta_R(0,0)\check f_R(0,0)\right\rangle e^{-i \psi(k_1,k_2)}\right) d(k_1,k_2)
\end{equation}
where we define the phase function 
\begin{equation}
\label{eq:phase}
\psi(k_1,k_2) = \langle \mathbf k-\mathbf k_r,\mathbf x_\Omega\rangle + 2 k_0 x_{M,3} + \left(\sqrt{k_0^2-k_1^2-k_2^2}- k_0\right) r_0 + \langle \mathbf k_r,\mathbf x_\mathcal D\rangle - k_0 r_\mathcal D
\end{equation}
and use $\mathcal B$ as given in \eqref{eq:int_domain}.

\section{Calibration of the Forward Model}
\label{sec:calibration}

So far we have investigated both the modeling of the backscattered wave from a (layered) object under Gaussian laser illumination and the measurement process of the OCT system described in Section~\ref{subsec:setup}. 

However, simulations based on this explicit model and the following comparison with experimental data presupposes the knowledge of a list of system parameters. Within this list we distinguish between parameters with values known from specifications such as the wavenumber $k_0$ and parameters which we need to calibrate from the experiment, the beam radius of the Gaussian beam and the angle of acceptance, for example. In order to be capable of extracting these parameters for the simulations, we use two calibration experiments. On the one hand, we consider an experiment which shows the behavior of the backreflected laser power for different surface tilting angles and on the other hand, we consider the influence of varying positions of the object, with respect to the focus, on the measured data.       

\subsection{Angular Dependence of the Measured Power}
\label{power_angle_section}
We use a mirror as a sample and analyze the influence of the surface angle on the measured intensity of the scattered electric field. Following the measurement process in Sections~ \ref{subsec:setup} and \ref{subsec:data_prep}, the reference arm is blocked, preventing any light from the reference arm to reach the detector. Furthermore, one of the two fibers, which would normally enter the DBD, is connected to a power meter. The measured data is therefore given as the intensity of the scattered field of this mirror. Hereby, again referring to the experimental setup in Section~\ref{subsec:data_prep}, we model the totally reflecting mirror as a sample characterized by an infinitely large refractive index. We parametrize the unit normal vector of the mirror surface by 
\begin{equation}
\label{eq:unit_normal}
\bm\nu_\Omega = \begin{pmatrix}\sin\theta_\Omega\\ 0\\\cos\theta_\Omega\end{pmatrix},
\end{equation}
for small values of $\theta_\Omega\in[\underline{\theta_\Omega},\, \overline{\theta_\Omega}].$

We describe the measurement process for this experiment in a way, that the scattered light is detected by a single scan lens point, for simplification we say $\mathbf x_\mathcal D = \mathbf 0,$ for a selected wavenumber $k_0$ in the spectrum $[k_{min},k_{max}].$ This in the end, yields a measured intensity of the form  
\begin{equation}
\label{eq:meas_angle}
\mathcal M_1(\theta_\Omega) =  |\tau \mathbf E_S^{(1)}(\mathbf 0)|^2,\quad \theta_\Omega\in[\underline{\theta_\Omega},\, \overline{\theta_\Omega}],
\end{equation}     
where $\mathbf E^{(1)}_S$ is given by \eqref{eq:samplefield} and $\tau\in\C$ accounts for the traveling through the beam splitters. Additionally, we say that the function $\check f_S$ is approximately given as in \eqref{eq:fexp} with $a = w_0^2/4,$ where $w_0$ represents the radius of the Gaussian beam at the focus. Following the experimental setup we fix the location $r_0 < 0$ (below the detector) of the focus and the mirror $x_{\Omega,3}$ and assume that they are equal: $r_0 = x_{\Omega,3}.$ We follow the notation from Section~\ref{subsec:near_far}, but approximate this time the exact form of the domain of integration $\mathcal B$ defined in \eqref{eq:int_domain}, which is an ellipse, by the rectangular domain  
\[
\mathcal B\approx [-L_1(\theta_\Omega) -k_0\sin(2\theta_\Omega), L_1(\theta_\Omega)-k_0 \sin(2\theta_\Omega)]\times [-L_2,L_2]
\]  
with the parameters  
\[
L_1 = k_0(-\sin(2\theta_\Omega-\theta_{max}) + \sin(2\theta_\Omega)),\ L_2 = k_0\sin(\theta_{max})
\]
and assume that this characterization of $\mathcal B$ still allows for an approximation of directions as in \eqref{eq:small_deviation}. Then, using the definitions of $\psi_j$ for $j\in\{0,1,2\}$ in \eqref{eq:parameters}, we obtain the intensity of the scattered field as a function of $\theta_\Omega,w_0$ and $\theta_{max}$ 
\begin{align}
\label{eq:scatt}
\left|\tau\mathbf E_S^{(1)}(\mathbf 0)\right|^2 =& G(\theta_\Omega,w_0,\theta_{max}),\\ \nonumber
G(\theta_\Omega,w_0,\theta_{max}) =& \frac{|\tau L_1(\theta_\Omega)|^2}{ 8 |\psi_2|^2 \pi^5 }
\left|\int_{\R}e^{- \frac{(\psi_1-\zeta)^2}{4\psi_2}} e^{i k_0 \sin(2\theta_\Omega) \zeta} \sinc(L_1(\theta_\Omega)\zeta) d \zeta \erfii\left(L_2\sqrt{\psi_2}\right)\right|^2,    
\end{align}
where $\sinc:\R\to\R$ denotes the unnormalized sinc function given by $\sinc(x)=\frac{\sin(x)}{x}$ and $\erfii:\C\to\C$ is the imaginary error function, defined by $\erfii(z) = \frac{2}{\sqrt{\pi}} \int_{0}^{z} e^{\zeta^2} d\zeta.$
Thus, from measurements $\mathcal M_1(\theta_\Omega)$ as in \eqref{eq:meas_angle}, corresponding to the data provided by our power meter, for different values $\theta_\Omega\in[\underline{\theta_\Omega},\, \overline{\theta_\Omega}],$ we can extract the beam radius $w_0$ at the focus and the angle of acceptance $\theta_{max}$ as solutions of the minimization problem
\[
(w_0,\theta_{max}) = \displaystyle \argmin_{(z_1,z_2)\in\R^2}\int_{\underline{\theta_\Omega}}^{\overline{\theta_\Omega}}\left| \mathcal M_1(\theta) - G(\theta,z_1,z_2) \right|^2 d \theta 
\]  
with the function $G$ given by \eqref{eq:scatt}.

\subsection{Reconstructing Sample Information from an OCT Experiment}
\label{sec:pow_foc}
In the previous section the beam radius $w_0$ at the focus and the angle of acceptance $\theta_{\text{max}}$ have been found. In order to complete the set of parameters necessary for the reconstruction from a measurement at a single point, we additionally need the normal vector $\bm\nu_\Omega$ of the tangential plane at each layer boundary.  

In swept source OCT one in-depth profile of the sample, that is a measurement of the form of $\mathcal M$ in \eqref{eq:dbd_data2} (in this case centered at $x_1=x_2=0$), called an A-scan, is acquired during one wavenumber sweep of the laser. To get 3D information, raster scanning in $x_1$ and $x_2$ direction over a certain field of view is performed. 
The data used in the following is considered as a B-scan, a line of A-scans where only $x_1$ varies at a fixed position $x_2$. Since we assume our layer boundaries to be planes with a certain normal vector $\bm\nu_\Omega,$ the surface points fulfill an equation of the form $\langle \mathbf x_\Omega,\bm\nu_\Omega\rangle = c.$ If we can therefore determine at every raster position $x_{\Omega,1},x_{\Omega,2}$ the third component $x_{\Omega,3},$ this determines the normal direction $\bm\nu_\Omega.$ 

Since the single A-scans along those lines are performed independently, we treat these A-scan as single measurements. We shift the coordinate system always so that the incident beam is located at $x_1=x_2=0$ and therefore have $\mathbf x_\Omega = x_{\Omega,3} \mathbf e_3.$ We recall, that the mirror in the reference arm is modeled as a medium described by an infinitely large refractive index with unit normal vector $\bm\nu_M = \mathbf e_3$ and fixed position at $\mathbf x_M = x_{M,3}.$ 

Under these assumptions, we rewrite \eqref{eq:cro} and \eqref{eq:phase} as 
\begin{equation}
\label{eq:cro_final}
\mathcal M_2(k_0) = -\frac{k_0 }{8 r_\mathcal D \pi^3} \int_{\mathcal B} \beta_{\Omega}(k_1,k_2)\check f_S(k_1,k_2)\check f_R(0,0) \sin\left( \psi(k_1,k_2)\right) d(k_1, k_2) 
\end{equation}
with $\beta_\Omega = \langle \bm \beta_S(k_1,k_2),\bm\beta_R(0,0)\rangle$ and
\[
\psi(k_1,k_2) = (k_3-k_{r,3})x_{\Omega,3} + 2 k_0 x_{M,3} + \left(\sqrt{k_0^2-k_1^2-k_2^2}- k_0\right) r_0 + (k_{r,3} - k_0) r_\mathcal D.
\]
For fixed mirror position $\mathbf x_M$, focus $r_0$ and detector $r_\mathcal D$, $\psi$ only varies with respect to different depth positions $x_{\Omega,3}$ of the the sample. Thus, if we can determine the function $\psi$ from the measurements $\mathcal M_2$ for different A-scans, we also obtain the depth information about the sample.      

Under the simplifying assumption that the far-field approximation of the scattered sample field, using $\mathbf s=\mathbf e_3$ in Theorem~\ref{thm:first_order}, is a reasonable approximation in this case, we rewrite \eqref{eq:cro_final} as 
\begin{equation}
\label{eq:cro_farfield}
\mathcal M_2(k_0) = \left(\frac{k_0}{2r_\mathcal D\pi}\right)^2 |c_3| \beta_{\Omega}(k_1,k_2)\check f_S(k_1,k_2)\check f_R(0,0) \cos\left(k_0 \frac{\psi(k_1,k_2)}{k_0}\right),
\end{equation}
where the point $(k_1,k_2)$ is defined by \eqref{eq:dir}. Since $\psi$ depends linearly on $k_0$, the measurements are then given as a harmonic oscillation with respect to $k_0$ and with frequency $\psi/k_0.$ To solve for this frequency, we want to Fourier transform with respect to $k_0$, which we define by
\[
\mathcal F(m)(\kappa) = \frac{1}{\sqrt{2\pi}}\int_\R m(k_0) e^{-i k_0 \kappa}d k_0.
\]
However, since we only have band-limited data, we will study the function 
\begin{equation}
\label{eq:intensity}
I:\R\to \R_+,\quad \kappa\mapsto \left|\frac{1}{\sqrt{2\pi}}\int_{k_{\text{min}}}^{k_{\text{max}}} \mathcal M_2(k_0) e^{-i k_0\kappa}d k_0\right|^2.
\end{equation}
We will show in the following that $\psi/k_0$ is determined as the argument where the maximum is located, that is, $\psi/k_0= \argmax_{\kappa} I(\kappa)$. (The absolute value is used to avoid real- or imaginary parts with higher frequent oscillations in order to stably calculate a maximal point.)

We assume that $\check f_S$ and $\check f_R$ in \eqref{eq:cro_farfield} are of exponential form as in \eqref{eq:fexp} and define the measurement function 
\[
M (\Theta_0;k_0) = \mathcal M_2(k_0) = K k_0^2 e^{-k_0^2 \sigma^2 }\cos(k_0 \Theta_0),
\]  
with the parameters 
\[
\sigma^2 = \frac{w_0^2}{4}\sin^2 (2\theta_\Omega), \quad  K = \frac{|c_3|}{(2\pi r_\mathcal D)^2} \beta_{\Omega}(k_1,k_2),\quad \Theta_0  = \frac{\psi}{k_0}.
\]
By rewriting
\[
M(\Theta_0;k_0)= -K \partial_{\sigma^2} e^{-k_0^2 \sigma^2 } \cos(k_0 \Theta_0)
\]
and interchanging the integral and differentiation in the Fourier transform $\mathcal F\left(\partial_{\sigma^2} e^{-k_0^2 \sigma^2 }\right),$ we find a form for the Fourier integral in \eqref{eq:intensity} as a convolution
\begin{multline}
\mathcal F(M)(\Theta_0;\kappa) = \frac{K \delta}{2\pi} \int_\R -\partial_{\sigma^2} \frac{1}{\sqrt{2\sigma^2}} e^{-\tfrac{1}{4\sigma^2} \zeta^2} \Big(\sinc(\delta(\kappa-\Theta_0-\zeta))e^{-i \bar k (\kappa-\Theta_0-\zeta) } \\
+ \sinc(\delta(\kappa+\Theta_0-\zeta)) e^{-i \bar k(\kappa+\Theta_0-\zeta) }\Big)d \zeta \label{eq:cross_fft}, 
\end{multline}
for $\bar k=\frac{k_{\text{max}}+k_{\text{min}}}{2}$ and $\delta=\frac{k_{\text{max}}-k_{\text{min}}}{2}.$ 
To simplify this expression, we introduce the values 
\begin{equation}
\label{eq:reference_values}
\sigma_{\bar k}=\frac{1}{\bar k},\quad \sigma_\delta=\frac{1}{\delta},
\end{equation} 
and observe from Table \ref{tab:tab2}, that $\sigma_{\bar k}$ and $\sigma$ are of the same order and $\sigma_\delta$ is considerably larger compared to both of them, meaning that
\begin{equation}
\label{eq:comparison}
\sigma_{\bar k} = Q \sigma, \quad\sigma \ll \sigma_\delta,
\end{equation}
for some $Q \in\R$ which is close to one.
Writing the functions under the integral \eqref{eq:cross_fft} in terms of these values gives us with
\begin{align}
u_{\sigma_{\bar k},\sigma}(\zeta) &= \frac{1}{(\sqrt{2}\sigma)^3}e^{-\tfrac{1}{Q^2}}\left(1-\frac{\zeta^2}{2\sigma^2}\right)e^{-\left(\frac{\zeta}{2\sigma} - i \frac{1}{Q}\right)^2},\label{eq:delta_function}\\
g_{\sigma_\delta,\pm}(\zeta) &=  \sinc\left(\frac{\zeta}{\sigma_\delta}\right)e^{-i \tfrac{1}{\sigma_{\bar k}} (\kappa \pm \Theta_0)}\nonumber. 
\end{align}
the expression
\begin{equation}
\label{eq:cross_fft1}
\mathcal F(M)(\Theta_0;\kappa) = \frac{K \delta}{2\pi} \int_\R u_{\sigma_{\bar k},\sigma}(\zeta) \left(g_{\sigma_\delta,-}(\kappa-\Theta_0-\zeta) + g_{\sigma_\delta,+}(\kappa+\Theta_0-\zeta)\right) d \zeta.
\end{equation}
Considering \eqref{eq:comparison}, we will expand this around $\sigma=0$.

\begin{lemma}
\label{lem:integral_approximation}
Let $\sigma_{\bar k},\sigma_\delta,\sigma$ as in \eqref{eq:reference_values} satisfying \eqref{eq:comparison}. Further, let $f_{\sigma_{\bar k},\sigma}$ be defined as in \eqref{eq:delta_function}. Then, we have for small values of $\sigma$ the approximation 
\begin{equation}
\label{eq:int_approx}
\int_\R u_{\sigma_{\bar k},\sigma}(\zeta)g_{\sigma_\delta,\pm}(\zeta) d\zeta \simeq \frac{1}{2\sqrt{2}}e^{-\tfrac{1}{Q^2}} \frac{4\sqrt{\pi}}{\sigma^2 Q^2} g_{\sigma_\delta,\pm}(0).
\end{equation}
\end{lemma}
Thus, by applying Lemma~\ref{lem:integral_approximation} to \eqref{eq:cross_fft1}, we obtain after changing back to the original system of coordinates
\[
\mathcal F(M)(\Theta_0;\kappa) \simeq \frac{  K \delta}{\sqrt{2\pi} }e^{-{\bar k}^2 \sigma^2} {\bar k}^2 \left( \sinc(\delta(\kappa-\Theta_0))e^{-i \bar k (\kappa+\Theta_0)} + \sinc(\delta(\kappa+\Theta_0))e^{-i \bar k (\kappa-\Theta_0)}\right),
\]
resulting in
\begin{multline}
\label{eq:sinc_app}
\left|\mathcal F(M)(\Theta_0;\kappa)\right|^2 \simeq \frac{  K^2 \delta^2}{2\pi }e^{-2 {\bar k}^2 \sigma^2} {\bar k}^4 \Big(\sinc(\delta(\kappa-\Theta_0))^2 \\
+ \sinc(\delta(\kappa+\Theta_0))^2 +2 \sinc(\delta(\kappa-\Theta_0))\sinc(\delta(\kappa+\Theta_0)) \cos(2\bar k\Theta_0) \Big). 
\end{multline}      
Note that the dominant sinc terms are centered symmetrically with respect to the origin. In order to derive an explicit expression for the maximum of \eqref{eq:sinc_app}, we want to assume that $\Theta_0$ is far away from the origin (which can be accomplished experimentally by tuning the position of the sample) then these sinc functions do not influence each other strongly. We shift one of them to the origin by setting $\kappa'=\kappa-\Theta_0$ and obtain 
\begin{multline}
\label{eq:sinc_app_shift}
F(\Theta_0;\kappa') \simeq \left|\mathcal F(M)(\Theta_0;\kappa'+\Theta_0)\right|^2\\
\simeq \sinc(\delta\kappa')^2 + \sinc(\delta(\kappa' + 2 \Theta_0))^2 +2 \sinc(\delta\kappa')\sinc(\delta(\kappa' + 2\Theta_0)) \cos(2\bar k\Theta_0).
\end{multline} 

\begin{lemma}
\label{lem:maximum}
Let $F$ be defined by \eqref{eq:sinc_app_shift}. Then, for $\Theta_0\to\infty,$ the function $F$ attains a local maximum at $\kappa' = 0.$ 
\end{lemma} 

Shifting back to the original coordinates and using Lemma~\ref{lem:maximum} yields that \eqref{eq:sinc_app} attains a maximum at $\kappa = \Theta_0,$ that is $\Theta_0 = \argmax_\kappa I(\kappa),$ which finally gives a representation of \eqref{eq:intensity} as
\begin{equation}
\label{eq:intensity2}
I(\Theta_0) \simeq \frac{  K^2 \delta^2}{2\pi }e^{-2 {\bar k}^2 \sigma^2} {\bar k}^4 \left(1 + \sinc(2\delta\Theta_0)^2 +2 \sinc\left(2\delta\Theta_0\right) \cos(2\bar k\Theta_0) \right).
\end{equation}
Thus, from the definition of $\Theta_0$ we can uniquely determine $\psi.$

We use this information for the reconstruction of the surface angle $\theta_\Omega.$ For two different, but known lateral positions $x^j_{\Omega,1},\ j\in\{1,2\},$ we consider A-scans leading to measurements $\mathcal M^j_2$ of the form \eqref{eq:cro_final}, for different depth positions $x_{\Omega,3}^j,$ for $j\in\{1,2\},$ respectively. By using the above analysis (under the assumption that the far-field approximation is valid), we determine from the Fourier transform of these two the phase contribution $\psi$ in dependence of $x_{\Omega,3}^1$ and $x_{\Omega,3}^2.$ Under the assumption that $\theta_\Omega$ is considered small, the subtraction of these two then leads to  
\[
\psi(x_{\Omega,3}^2)-\psi(x_{\Omega,3}^1) \simeq 2 ( x^2_{\Omega,3} - x^1_{\Omega,3} ),
\]   
which gives the difference in depth $( x^2_{\Omega,3} - x^1_{\Omega,3} ). $ Together with known lateral information and using that the unit normal vector on the surface satisfies $\langle \bm\nu_\Omega, (\mathbf x^2_\Omega - \mathbf x^1_{\Omega})\rangle = 0,$ we determine $\theta_\Omega$ as
\[\theta_\Omega = \arctan\left(\frac{x^2_{\Omega,1} - x^1_{\Omega,1} }{x^1_{\Omega,3} - x^2_{\Omega,3} } \right). \]

\section{Results}\label{sec:results}

Finally, we want to validate our model by comparing the simulations with experimental data. We focus on the two previously addressed experiments (see Sections~\ref{power_angle_section} and \ref{sec:pow_foc}, respectively \ref{subsubsec:data_angle} and \ref{subsubsec:data_focus}). This quantitative approach shows the dependence of the data on the surface tilting and the focus position. 

First we will use the calibration measurement to calculate the beam radius at the focus and the angle of acceptance, then we look (using the just calibrated parameters) at the data from Section~\ref{subsubsec:data_focus}. The main parameters are presented in Table~\ref{tab:tab2}.

\subsection{Power vs. Angle - Experiment}\label{subsec:results_angle}

Following Section~\ref{power_angle_section}, we first calibrate the beam radius and the angle of acceptance from the experimental data for different values of the surface angle. This procedure is presented in Algorithm~\ref{alg1}. The algorithm is based on the approximated form \eqref{eq:scatt} shortening the computation time by evaluating a one-dimensional integral, instead of the two-dimensional integration presented in \eqref{eq:samplefield}. As discussed in Section~\ref{subsubsec:data_angle}, the sample arm power arriving at one of the DBD entrances was measured $M=6$ times at $J=25$ angular positions.  
While the laser power behaves very stable, due to the expected error from the rotational stage and the Gaussian dependence from the angle, some error is observed in the power vs. angle data. Thus, in the following, the data will be plotted with errorbars, representing the standard deviation.

\begin{algorithm}[hbt!]
 \caption{Extraction method for the beam radius $w_0$ and the angle $\theta_{max}$ of acceptance.}\label{alg1}

\KwResult{ $w_0$ and $\theta_{\text{max}}$  }
\textbf{Input:} wavenumber $k_0=\frac{2\pi}{\lambda_0}$ with the central wavelength $\lambda_0=\SI{1300}{\nano\metre}$, $\mathcal M_1(\theta_{j})$, for $j=1,\dots,J$ ; \\
$(w_0,\theta_{max}) = \argmin_{(z_1,z_2)} \frac{1}{J}\sum_{j=1}^J \left(\mathcal M_1(\theta_j) - G(\theta_j,z_1,z_2) \right)^2;$\\[1ex]
\end{algorithm}

\begin{figure}[ht]
\centering
\includegraphics[width=0.65\textwidth]{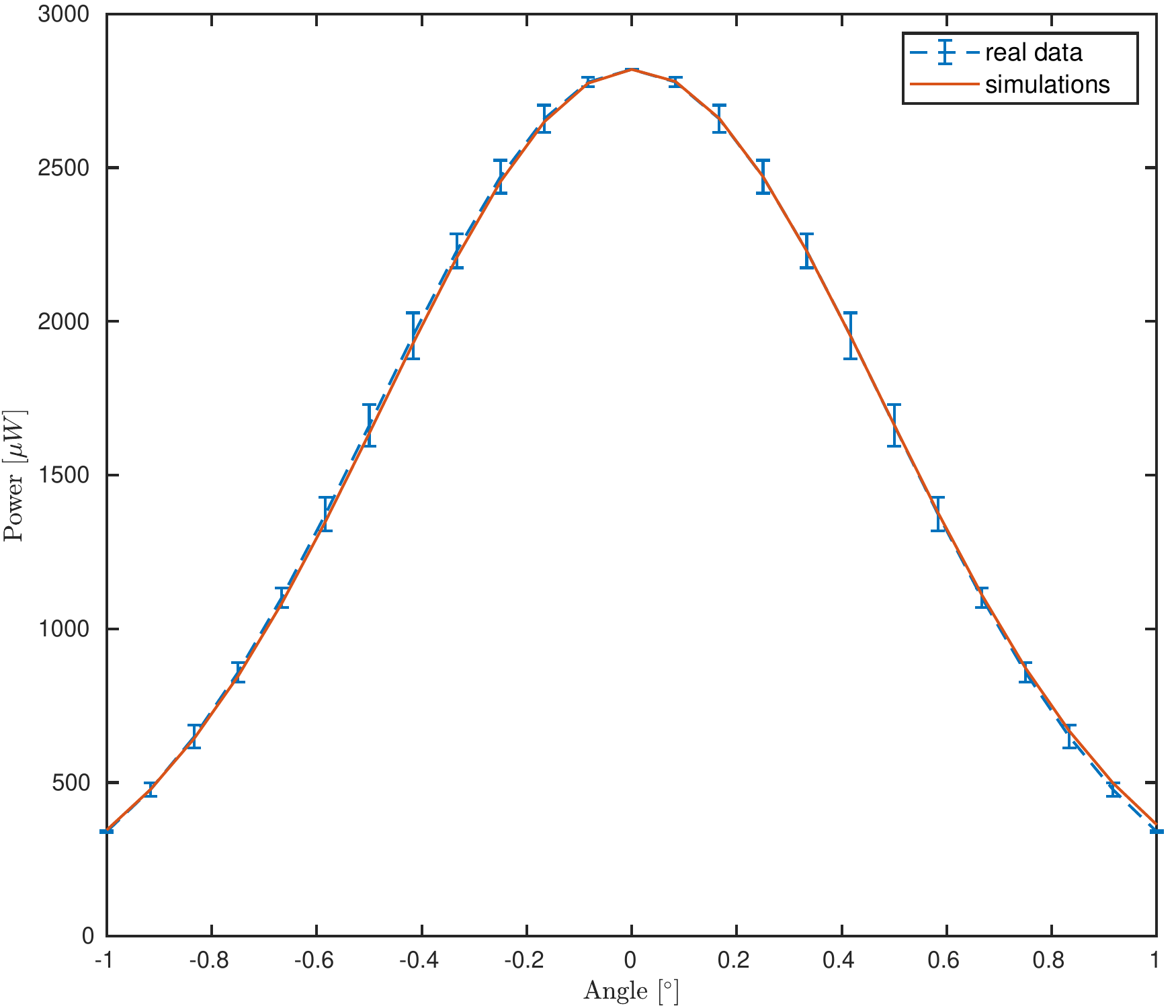}
\caption{Comparison between the power meter measurements for different angular steps of the mirror (blue dashed curve) as in Section~\ref{subsec:data_prep} and the simulation (red curve) for \eqref{eq:scatt}.} \label{fig:angle_power}
\end{figure}

The simulated data $\mathcal{M}_1,$ see \eqref{eq:meas_angle}, is given for $\theta_\Omega = \theta_j, \, j = 1,...,J$. The match between experimental and simulated data is presented in Figure~\ref{fig:angle_power}. We remark that both data and simulation follow a Gaussian behavior and attain the maximum at normal incidence, as expected.

\begin{table}
\centering
\begin{tabular}{m{6cm} m{4cm} m{1.5cm}}
 Parameter & Value & Unit\\ \hline
 $\lambda_0$ central wavelength  &  $1300$  &  \si{\nano\metre}   \\ 
 $w_0$  beam radius at focus &  $14.15$  &  \si{\micro\metre}   \\
 $\theta_{max}$ angle of acceptance & 1.5709 & \si{\degree} \\
 $NA$	& 0.037 & \\
 $k_{\text{min}}$ 	&    $4.7835$ & \si{\per\micro\metre} \\
 $k_{\text{max}}$ 	&    $4.8973$ & \si{\per\micro\metre} \\
 $\bar k=\frac12(k_{\text{max}}+k_{\text{min}})$ 	&    $4.8404$ & \si{\per\micro\metre} \\
 $\delta=\frac12(k_{\text{max}}-k_{\text{min}})$ 	&    $0.056858$ & \si{\per\micro\metre} \\
 $\sigma=\frac{w_0}2\left|\sin(2\theta_\Omega)\right|$, $\sigma_{\bar k}=\bar k^{-1}$, $\sigma_\delta=\delta^{-1}$ 	&  $0.2753$, $0.2066$, $17.588$  & \si{\micro\metre} \\
\end{tabular}
\caption{List of parameters: The central wavelength and the wavenumbers $k_{\text{min}}$ and $k_{\text{max}}$ are determined through the specifications of the used swept-source. The beam radius and the angle of acceptance were found through calibration (see Section~\ref{subsec:results_angle}). The numerical aperture NA was calculated from the angle of acceptance. The parameter $\sigma$ is given for a typical tilting angle $\theta_\Omega$ in our experiments. \label{tab:tab2}}
\end{table}
  
\subsection{Power vs. Focus - Experiment}
Using the calibrated spot size and angle of acceptance from the previous experiment, we compare the simulations with experimental data for multiple B-scans of a mirror and a coverglass as samples of interest.

Following Section~\ref{sec:pow_foc}, we first determine the surface angle $\theta_\Omega$ and adapt the integration area in \eqref{eq:int_domain}. The coverglass, which is described by a medium with constant refractive index $n_1 = 1.5088$ and perfectly parallel surfaces, has a thickness $d$, which is also determined from the experimental data, see Algorithm~\ref{alg2}. 

The experimental data is measured at a series of different wavelengths $\lambda_j\in (\SI{1282.86}{\nano\metre},\SI{1313.71}{\nano\metre})$, $j=1,\dots,J,\ J=700,$ equally spaced in wavenumber $k_{0,j}=\frac{2\pi}{\lambda_j}.$ As described in Section~\ref{subsubsec:data_focus}, the sample was imaged at different positions $x_{\Omega,3}=x_{n},\, n=1,\dots,N,$ along the depth axis.

We ignore polarization effects in the following and use the form \eqref{eq:samplefield} with $\beta_S = 1$ for the simulations of the scattered field of the mirror data. However, for the coverglass experiment we extend the form to a layer model with two parallel surfaces
\begin{align*}
\mathbf E^{(1)}_S(\mathbf x) = \frac{1}{4\pi^2} \int_{\mathcal B} \beta_S(k_1,k_2) \check f_S(k_1,k_2) e^{-i\sqrt{k_0^2-k_1^2-k_2^2}r_0} e^{-i \langle(\mathbf k-\mathbf k_r), \mathbf x_\Omega\rangle} e^{-i \langle\mathbf k_r, \mathbf x\rangle} d (k_1,k_2),
\end{align*}
where we have given the reflection coefficient 
\[
\beta_S(k_{1},k_{2}) = \beta_{0} -  \beta_{0}(1-\beta^2_{0})e^{-i k_0 2 n_1 d \cos\theta_t}.
\]
We assume again that $\check f_S$ can be well approximated by \eqref{eq:fexp} and define $\mathcal B$ as in \eqref{eq:int_domain}.

\begin{algorithm}[hbt!]
 \caption{Extraction scheme for the tilting angle $\theta_\Omega$ and the thickness $d$ from the power vs. focus experiment.}\label{alg2}
\KwResult{$\theta_\Omega$ and $d$ }
\textbf{Input:} refractive index $n_1$, lateral pixel size size $P_x,\, k_{0,j},$ for $j=1,\dots,J,$ \\
B-scan $\mathcal S=\{\underline{N},\dots,\bar{N}\}\subset\{1,\dots,N\},\, |\mathcal B| = \tilde N;$ \\
$\mathcal M_2(x_n;k_{0,j}) $, for $n\in\mathcal S,\ j=1,\dots,J$ ; \\
 \tcc{Extraction of the surface angle $\theta_\Omega$ and the thickness $d.$}
\For {$\underline N\leq n\leq \bar N$}{
$ \mathcal{F}M_{n,l}=\frac{1}{\sqrt{2\pi}}\sum_{j=1}^J \mathcal M_2(x_n;k_{0,j}) e^{-i k_{0,j} z_l},\, l=1,\dots,\tilde L;$\\
$p_{n,1} = \argmax_{l}\left|\mathcal{F}M_{n,l}\right|;$\\ 
$p_{n,2} = \argmax_{\left\{l\ |\ z_l > p_{n,1} + \tfrac{4\pi}{k_{0,J}-k_{0,1}} \right\}} \left|\mathcal{F}M_{n,l}\right|;$ 
}
$\theta_\Omega = \arctan\left(2\frac{\tilde N \cdot P_x}{p_{\underline N,1}-p_{\bar N,1}}\right);\quad \cos\theta_t = \sqrt{1-\frac{1}{n_1^2}\sin^2(\theta_\Omega)}; $\\
$\tilde{x}_{1} = (0,0,p_{\underline N,1}/2); \quad \tilde{x}_{\tilde N} = (\tilde N \cdot P_x,0,p_{\bar N,1}/2); \quad d = \frac{1}{\tilde N}\sum_{l=\underline N}^{\bar N} \frac{p_{l,2}-p_{l,1}}{2 n_1\cos\theta_t};$\\
\end{algorithm}

Although the dual-balance detection already lowers the noise level in the data, we need to minimize the effects of the residual noise for a meaningful comparison with the simulations.

\begin{figure}[hbt!]
\centering
\includegraphics[width=0.65\textwidth]{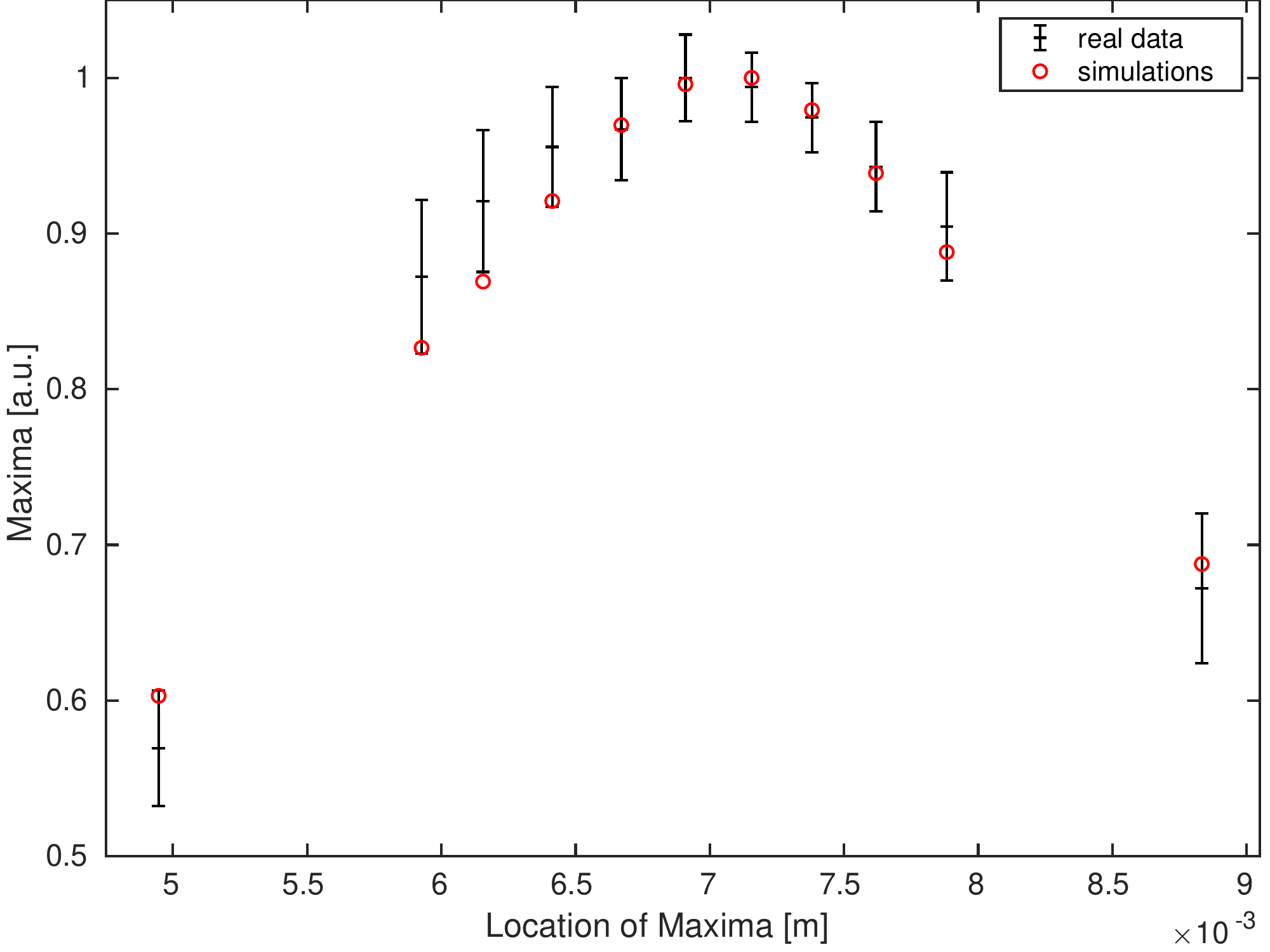}
\caption{Comparison of averaged maximum values for all B-scans (of different sample locations) of the experimental data (black with errorbars) and the simulated data points (red) for the mirror experiment. } \label{fig:focus_power1}
\end{figure}

Thus, for every individual B-scan, we use the mean value of all maximum intensities of it's A-scans, i.e. we consider the map 
$
\frac{1}{\tilde N}\sum_{n=1}^{\tilde N} \max_{\kappa\in\R}I_n(\kappa),
$
where we use $I_n = I,$ as defined in \eqref{eq:intensity}, with $n\in\{1,\dots,\tilde N\}$ accounting for the number of A-scans in every B-scan and for the different position $x_{\Omega,3}^n$  (corresponding to this A-scan), as a reference for our simulations. 
The errorbars in Figure~\ref{fig:focus_power1} represent the standard deviation of these variations over each B-scan. The highest value of the experimental data is matched to the highest value of the simulation for comparison.

Figure~\ref{fig:focus_power1} shows, that the Gaussian behavior of the data for the mirror experiment follows the theory.

\begin{figure}[ht]
\centering
\includegraphics[width=0.65\textwidth]{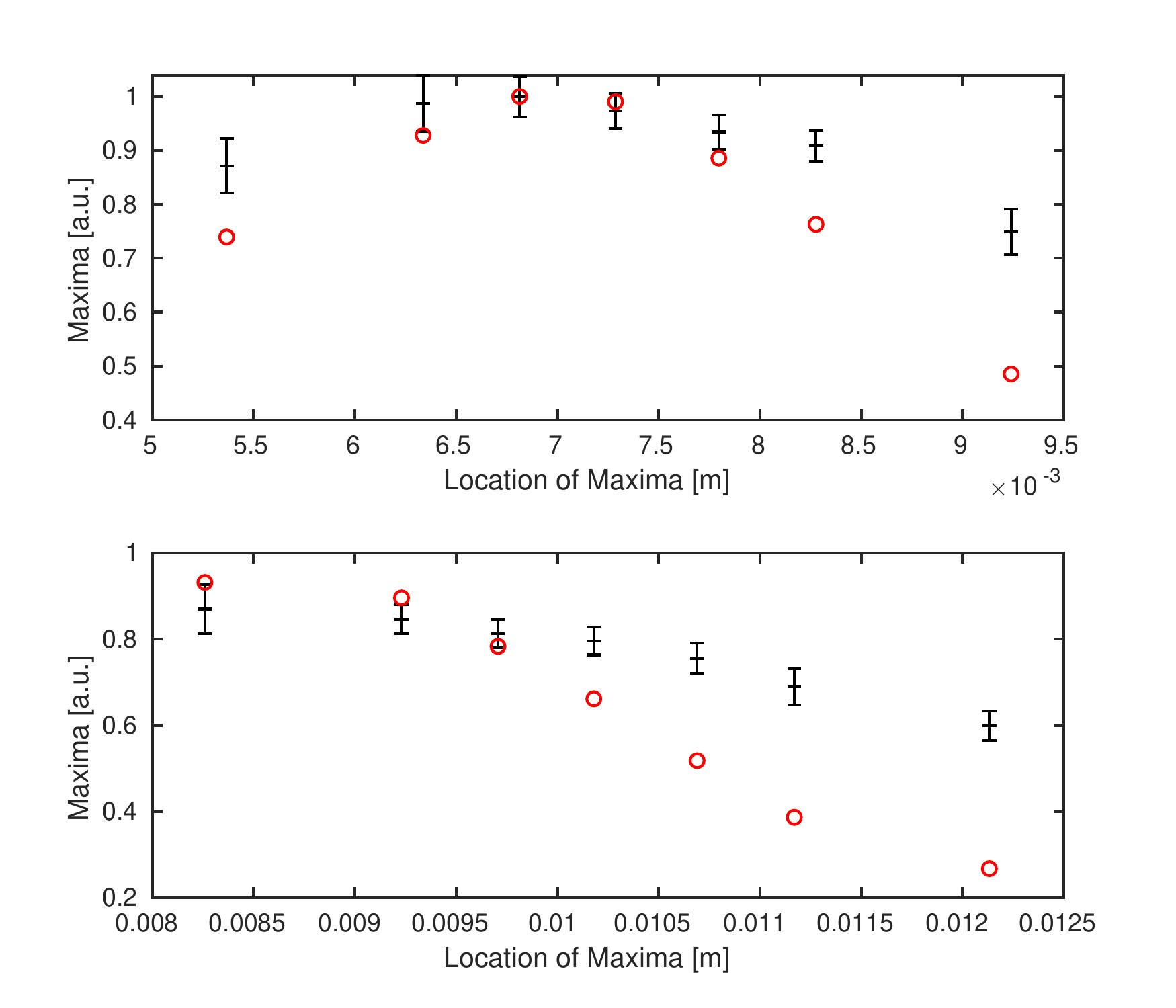}
\caption{Comparison between the experimental data (black with errorbars) and the simulated (red) data points for calibrated values of $w_0$ and $\theta_{max}$. Above we see the top boundary surface of the coverglass, below the background surface.} \label{fig:focus_power3}
\end{figure}

In Figure~\ref{fig:focus_power3}, we see the comparison between experimental data and simulations for both boundaries of the coverglass.
Unfortunately, the calibrated parameters $(w_0,\theta_{max})$ from the previous subsection do not yield optimal results, see Figure~\ref{fig:focus_power3}. Similar to Figure~\ref{fig:focus_power1} a sufficiently strong decrease of the averaged maximum values away from the focus position can be identified in the simulations for the coverglass experiment as well.        

\begin{figure}[hbt!]
\centering
\includegraphics[width=0.65\textwidth]{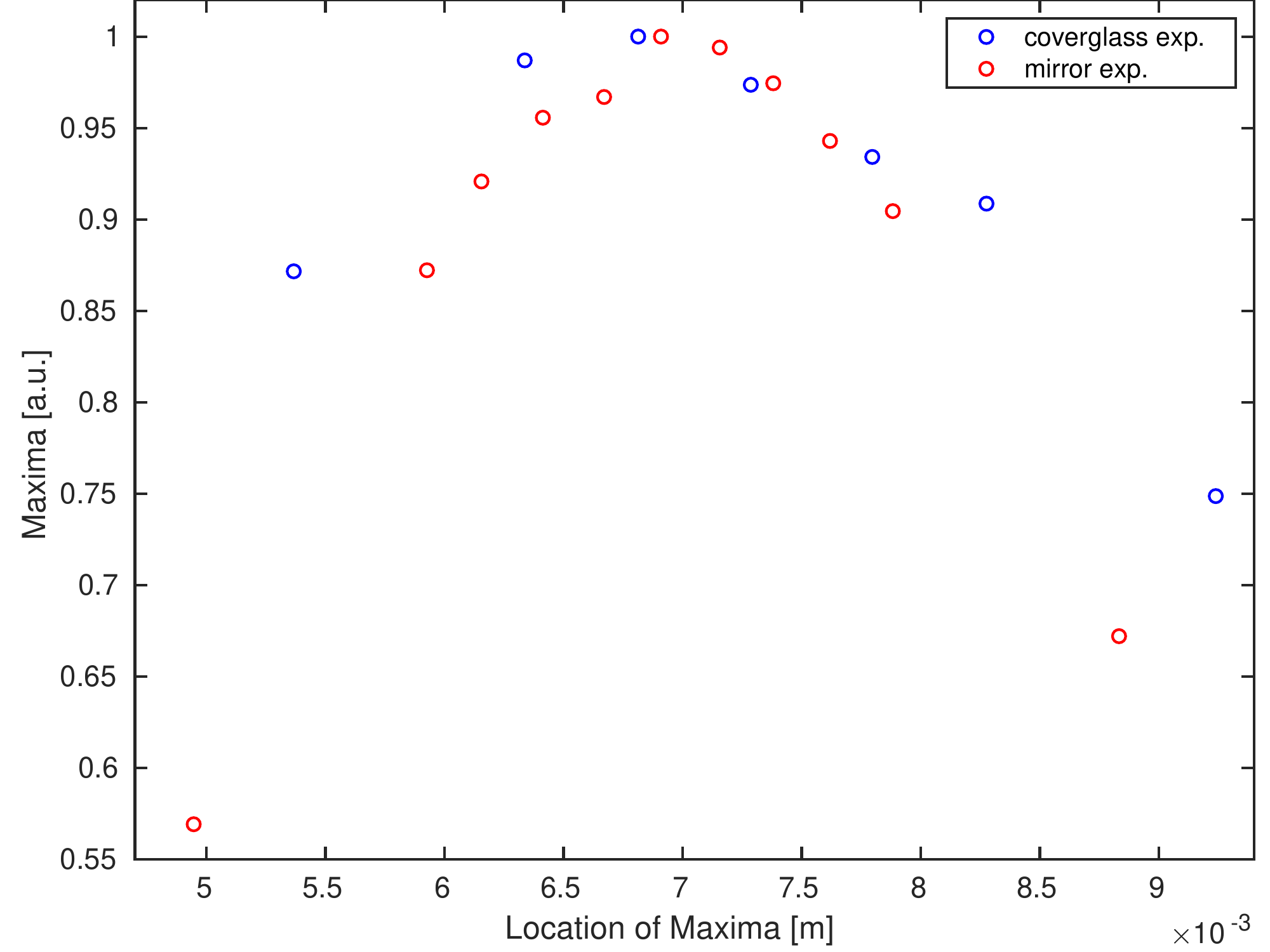}
\caption{Comparison between averaged maximum values for all B-scans of the mirror (red) and the coverglass (blue) experiment. } \label{fig:gauss_behav}
\end{figure}

At this point, we remark by comparing the experimental data sets, see Figure~\ref{fig:gauss_behav}, that the Gaussian curve for the coverglass experiment shows a slightly stretched behavior. This is explained by the measurement of the returning laser light in a diffusive regime originating from the reflection at a slightly rough coverglass boundary surface.
 
\begin{figure}[ht]
\centering
\includegraphics[width=0.65\textwidth]{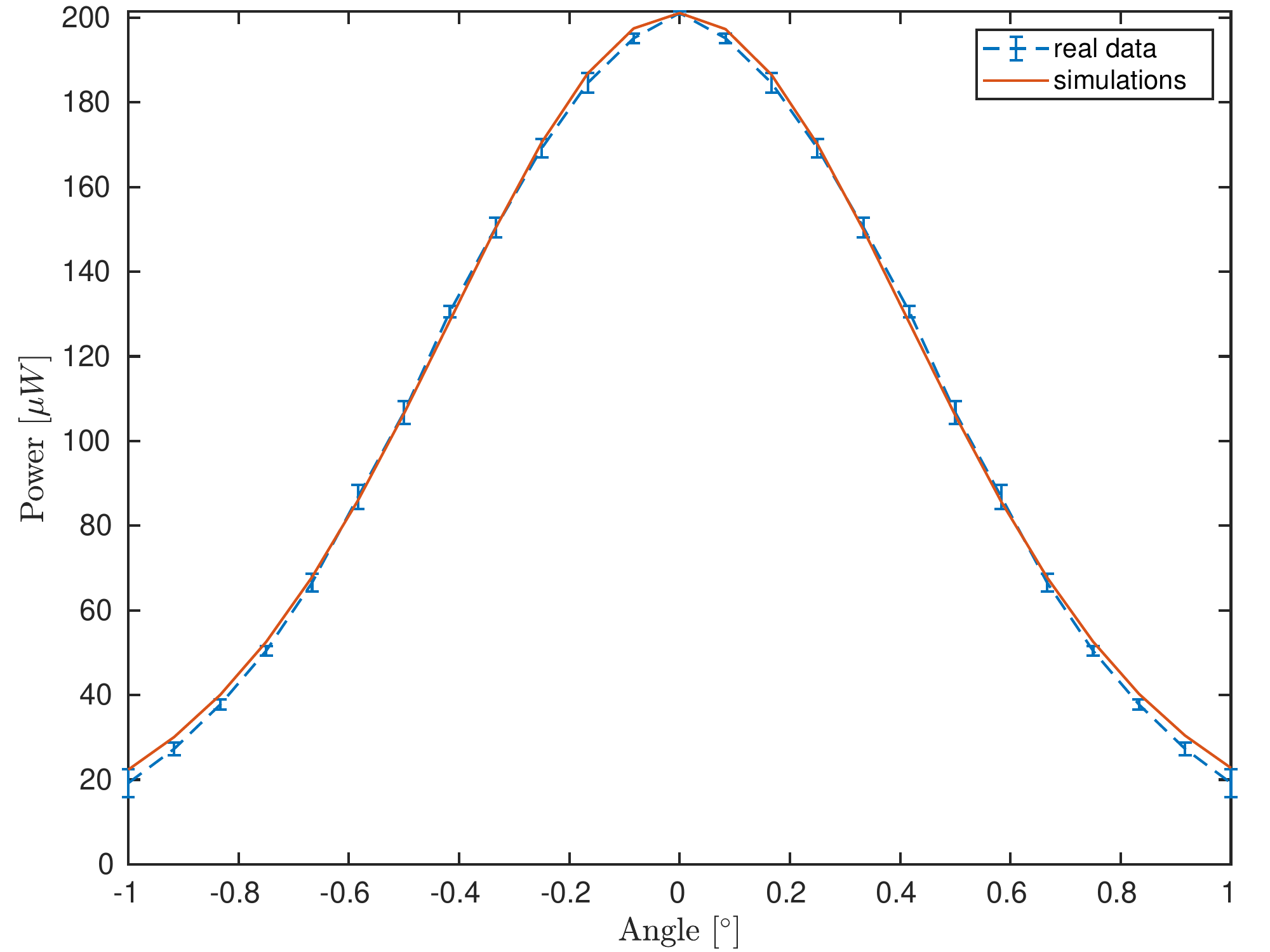}
\caption{Angular scattering profile of a coverglass: comparing experimental data (blue dashed line) with simulations (red).} \label{fig:angle_power2}
\end{figure}

\begin{figure}[ht]
\centering
\includegraphics[width=0.65\textwidth]{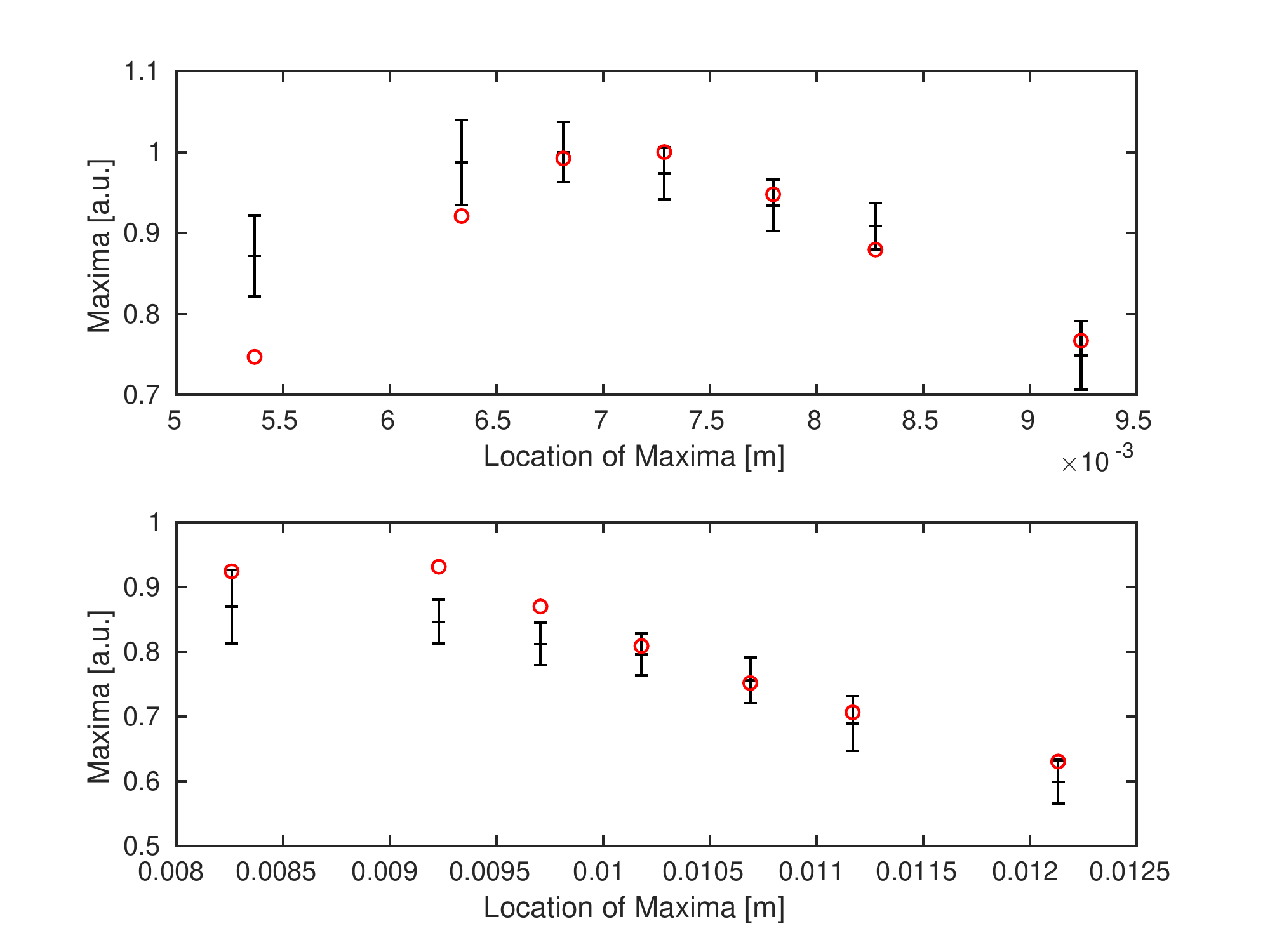}
\caption{Comparison between averaged maximum values of the experimental data (black) and simulations (red) for different position through the focus after the recalibration of values $w_0$ and $\theta_{max}$. Above we see the top surface, below the background surface of the underlying coverglass.} \label{fig:power_focus}
\end{figure}

However, updated parameters can be found using an experiment similar to the calibration of $w_0$ and $\theta_{max}$, see Figure~\ref{fig:angle_power2}. In contrast to the calibration, the power measurement for the coverglass includes information of both boundaries and therefore the measured field intensity is provided as a sum  
\[
\left|\mathbf E_S^{(1)}\right|^2 = \left|\mathbf E^{(1)}_{S,1} + \mathbf E^{(1)}_{S,2}\right|^2,
\]
where we consider first order reflections only. Due to additional scattering events inside the coverglass material, the background information $\mathbf E^{(1)}_{S,2}$ is smaller than $\mathbf E^{(1)}_{S,1}$ and therefore neglected for the calibration.
A comparison for the coverglass experiment -- after updating the system parameters -- shows that almost all simulated data points lie inside the estimated range for both boundaries, see Figure~\ref{fig:power_focus}.     

\section{Discussion}
We have considered samples with a very distinct scattering profile, but still a slight difference in the angular reflectivity profile could be observed for the mirror and the coverglass. For diffusely scattering samples the proposed description can easily be generalized, especially with the mentioned automatic angular power measurement. An automated measurement would also reduce the error, so that a fast angle scanning procedure could be implemented prior to OCT imaging.

The error in the power vs.\ focus experiment (see Figures \ref{fig:focus_power1} and \ref{fig:focus_power3}) is caused by power variations inside single B-scans. These are explained by a combination of reasons. The tilt of the sample with respect to the OCT illumination, generates a slight continuous change in distance to the focus for each A-scan inside a B-scan. In addition the scan lens induces a certain curvature of field, resulting in a change of illumination depending on the raster scanning position $x_1$.

Although the presented results in this work show suitable correspondence with the provided OCT data, we note that the algorithm for solving the minimal-error-solution problem in extracting the beam radius and the angle of acceptance from the power-angle experiment suffers from the fact that the function in \eqref{eq:scatt} is highly oscillating and it is therefore difficult to find an ``optimal'' set of parameters.

Nevertheless, for a rather large range of values of these parameters, we get a reasonable match with the experimental data, at least for simple, layered examples. The model should, however, work nicely also for more complicated samples with different geometries (that is, with more and potentially curved layers).

We expect that this model can be used as a forward model of the inverse problem of reconstructing the refractive index inside of the layers.

\section{Conclusion}
We presented a method to model the image formation in OCT based on a real-life \SI{1300}{\nano\metre} swept-source setup. In contrast to publications based on plane wave models for the OCT system, the proposed model includes the effect of additional system relevant parameters such as the focus and the beam radius of the incident laser light and the angle of acceptance. We also suggested a way how to determine these (not necessarily a priori known) parameters, either from the OCT data itself and from calibration measurements.

A comparison between simulation and experiment shows, that the presented model, together with the derived system and sample parameters, produces a quantitatively correct prediction of the OCT data. We therefore expect that this model can serve as a forward model for an inversion algorithm to quantitatively reconstruct from OCT data the optical material properties of the sample, in particular its refractive index.


\section{Appendix}
Here, we collect the proofs of the theorems. We start with the one of Theorem~\ref{thm:thm1} which describes the Fourier decomposition of a solution of the Helmholtz equation.

\begin{proof}[Proof of Theorem~\ref{thm:thm1}]
To simplify the notation, we shift the coordinate system $x_1'=x_1,x_2'=x_2,x_3'=r_0-x_3$ such that focal plane is in the origin of the new system. Then, we apply the Fourier transform in the first equation of \eqref{eq:Helmholtz_system} with respect to the $x_1,x_2$-components, resulting in an ordinary differential equation 
\[
\partial_{x_3}^2 \check E_j(k_1,k_2,x_3')+(k_0^2-k_1^{2}-k_2^{2})\check E_j(k_1,k_2,x_3')=0, 
\]
for the first two components of the electric field. We know that for $j\in\{1,2\},$
\[
\check E_j(k_1,k_2,x_3')=\alpha_{-,j}(k_1,k_2) e^{- i \sqrt{k_0^2-(k_1^{2}+k_2^{2})}x_3'} + \alpha_{+,j}(k_1,k_2) e^{ i \sqrt{k_0^2-(k_1^{2}+k_2^{2})}x_3'}
\]
is a solution of this problem. Using the Fourier transformed initial data at the plane $x_3'=0,$ we find that the coefficients $\alpha_{-,j},\alpha_{+,j}$ are given by
\[
\alpha_{-,j}(k_1,k_2) + \alpha_{+,j}(k_1,k_2) = \check f(k_1,k_2)p_j, \quad j\in\{1,2\}.
\]
So far we have seen that  $E_1,E_2$ are solutions of the Helmholtz equation, without considering the third component of $\mathbf E.$ Finally, we use that $\mathbf E$ is divergence-free to find that
\begin{equation}
\label{eq:third_component}
E_3(\mathbf x')=-\int_{0}^{x_3'}\left(\partial_{x_1} E_1(x_1',x_2',z) + \partial_{x_2} E_2(x_1',x_2',z)\right)d z.
\end{equation}
Moreover, taking two times the derivative with respect with $x_3,$ we find
\begin{align*}
\partial^2_{x_3} E_3(\mathbf x') &= -(\partial_{x_1} \partial_{x_3} E_1(x_1',x_2',x_3')+\partial_{x_2} \partial_{x_3} E_2(x_1',x_2',x_3'))\\
 &=-\int_{0}^{x_3'}\left( \partial_{x_1} \partial^2_{x_3} E_1(x_1',x_2',z) + \partial_{x_2} \partial^2_{x_3} E_2(x_1',x_2',z)\right) d z \\
 &+ \partial_{x_1} \partial_{x_3} E_1(x_1',x_2',0) + \partial_{x_2} \partial_{x_3} E_2(x_1',x_2',0),
\end{align*}
which in the end, using also the second derivatives with respect to $x_1$ and $x_2$, gives 
\[\Delta E_3(\mathbf x')+k_0^2 E_3(\mathbf x') =\partial_{x_1} \partial_{x_3} E_1(x_1',x_2',0) + \partial_{x_2} \partial_{x_3} E_2(x_1',x_2',0). \]
Thus, $E_3$ is also solution of the  Helmholtz equation if and only if
\[
\partial_{x_1} \partial_{x_3} E_1(x_1',x_2',0) + \partial_{x_2} \partial_{x_3} E_2(x_1',x_2',0)=0.
\]  
This is equivalent to the condition
\[k'_1 (-\alpha_{-,1}(k_1,k_2) + \alpha_{+,1}(k_1,k_2)) + k_2(-\alpha_{-,2}(k_1,k_2) + \alpha_{+,2}(k_1,k_2)) =0. \]
Since this condition must hold true for every $(k'_1,k'_2)\in\R^2,$ we get  
\[
\alpha_{-,j}(k_1,k_2)=\alpha_{+,j}(k_1,k_2), \quad j\in\{1,2\}
\]
and therefore 
\[
\alpha_{-,j}(k_1,k_2)=\alpha_{+,j}(k_1,k_2)=\frac{1}{2}\check f(k_1,k_2) p_j, \quad j\in\{1,2\}. 
\] 
Given the representations of $E_j,\,j\in\{1,2\},$ we derive a representation also for the third component of the electric field
\begin{align*}
E_3(\mathbf x') &=\frac{1}{8\pi^2} \int_{\R^2}\left( -\check f(k_1,k_2)\frac{p_1 k_1 + p_2 k_2}{\sqrt{k_0^2-(k_1^{2}+k_2^{2})}} e^{-i (k_1 x_1' + k_2 x_2')} e^{- i \sqrt{k_0^2-(k_1^{2}+k_2^{2})}x_3'}\right.\\
&\left.+ \check f(k_1,k_2) \frac{p_1 k_1+ p_2 k_2}{\sqrt{k_0^2-(k_1^{2}+k_2^{2})}} e^{-i (k_1 x_1' + k_2 x_2')}e^{ i \sqrt{k_0^2-(k_1^{2}+k_2^{2})}x_3'}\right) d (k_1,k_2).
\end{align*}
Finally, we use the original coordinate system and we obtain the desired representations \eqref{eq:incident_gaussian} and  \eqref{eq:initial_dist}.  
\end{proof}

Next, we come to the derivation of the far-field representation of the scattered field presented in Theorem~\ref{thm:first_order}. This is described, for example, in the book \cite{ManWol95} and is based on the stationary phase method.

\begin{lemma}
\label{lem:stationary_phase}
Let $\mathcal G$ denote the set of critical points of the function $\Psi:\R^2\to\R$ and assume that $u:\R^2\to\R^3$ is compactly support. Further assume that for every $\xi\in\mathcal G$ the Hessian matrix $H$ of $\phi$ satisfies
\[\det\left(H(\Psi)(\xi)\right)\neq 0.\] 
Then, we have asymptotically as $N\to \infty$ that 
\begin{equation*}
\int_{\R^2}u(x)e^{-i N \Psi(x)}d x \\
=e^{-i N \Psi(\xi)} \frac{1}{\sqrt{\det\left(\frac{N}{-2\pi i}H(\Psi)(\xi)\right)}} \sum_{\xi\in\mathcal G}  u(\xi)+ o(1/N).
\end{equation*}
\end{lemma}    
\begin{proof}
See \cite{Hoe03}[Theorem 7.7.5]. 
\end{proof}

We can now apply this stationary phase method, Lemma~\ref{lem:stationary_phase}, to the integral in \eqref{eq:samplefield_near}.

\begin{proof}[Proof of Theorem~\ref{thm:first_order}]
Considering in \eqref{eq:samplefield_near} with $\mathbf x=r\mathbf s$ the limit $r\to\infty$, we correspondingly define the phase function 
\begin{equation*}
\Psi(k_1,k_2)=\frac1{k_0}\langle\mathbf k_r,\mathbf s\rangle=\langle \tfrac{\mathbf k}{k_0}, \mathbf s\rangle - 2 \langle \tfrac{\mathbf k}{k_0 }, \nu_\Omega\rangle \langle \nu_\Omega, \mathbf s\rangle.
\end{equation*}
In order to calculate the critical points of $\Psi$, we look for solutions of the equation $\nabla \Psi(k'_1,k'_2) = 0.$ This gives us for the critical points the condition
\begin{equation}
\label{eq:critical_points}
\frac{s_j}{k_0} + \frac{ k_j s_3}{k_0\sqrt{k_0^2-k^{2}_1-k^{2}_2}} -2 \langle \nu_\Omega, \mathbf s\rangle \left( \frac{\nu_{\Omega,j}}{k_0} + \frac{ k_j \nu_{\Omega,3}}{k_0\sqrt{k_0^2-k^{2}_1-k^{2}_2} }\right)=0,
\end{equation}
for $j\in\{1,2\}.$ For the sake of simplicity, we define the parameters 
\[
c_j := s_j - 2\langle \nu_\Omega, \mathbf s \rangle \nu_{\Omega,j}, \quad j\in\{1,2,3\}, 
\]
that satisfy 
\begin{equation}
\label{eq:identity}
\sum_{j=1}^3 c_j^2 = 1.
\end{equation}
Now, rewriting \eqref{eq:critical_points}, we get 
\[
c_j = - \frac{1}{\sqrt{k_0^2-k^{2}_1-k^{2}_2}} k_j c_3.
\]
Then, the condition \eqref{eq:identity} implies that  
\begin{equation*}
\begin{pmatrix}k_1 \\ k_2 \end{pmatrix} = -k_0\sign(c_3)\begin{pmatrix} s_1 - 2\langle \nu_\Omega, \mathbf s \rangle \nu_{\Omega,1}\\ s_2 - 2\langle \nu_\Omega, \mathbf s \rangle \nu_{\Omega,2} \end{pmatrix},
\end{equation*} 
which is \eqref{eq:dir}. To show that the Hessian matrix
\[
H(\Psi)(k_1,k_2)=\begin{pmatrix}\partial^2_{k_1}\Psi(k_1,k_2) && \partial_{k_1}\partial_{k_2}\Psi(k_1,k_2)\\ \partial_{k_2}\partial_{k_1}\Psi(k_1,k_2) && \partial^2_{k_2}\Psi(k_1,k_2)  \end{pmatrix},
\]
is invertible at this position $(k_1,k_2)$,  we compute for $j,l\in\{1,2\}:$
\begin{align*}
\partial_{k_l}\partial_{k_j}\Psi(k_1,k_2) &= \frac{s_3-2\langle \nu_\Omega, \mathbf s \rangle \nu_{\Omega,3}}{ k_0\left(k_0^2-k_1^{2}-k_2^{2}\right)}\frac{k_0^2-k^{2}_1-k^{2}_2 + k_j^{2}}{\sqrt{ k_0^2-k^{2}_1-k^{2}_2}}, \quad l=j, \\
\partial_{k_l}\partial_{k_j}\Psi(k_1,k_2) &= \frac{s_3-2\langle \nu_\Omega, \mathbf s \rangle \nu_{\Omega,3}}{k_0\left(k_0^2-k_1^{2}-k_2^{2}\right)}\frac{k_j k_l}{\sqrt{ k_0^2-k^{2}_1-k^{2}_2}}, \quad l\neq j.
\end{align*}
Thus, the determinant of the Hessian matrix is given by 
\[
\det\left(H(\Psi)(k_1,k_2)\right) = \frac{1} { \left( s_3-2 \langle \nu_\Omega, \mathbf s \rangle \nu_{\Omega,3} \right) ^2 k_0^4} > 0,
\]
so that a direct application of Lemma \ref{lem:stationary_phase} to the integral in \eqref{eq:samplefield_near} gives us
\[
\mathbf E_{S,\infty}(r\mathbf s) = \frac{ -i k_0 \left| c_3 \right|}{2\pi r} \bm\beta_S(k_1,k_2)\check f_S(k_1,k_2) e^{-i\sqrt{k_0^2-k_1^2-k_2^2}r_0}e^{-i\langle \mathbf k-\mathbf k_r,\mathbf x_\Omega\rangle}  e^{i k_0 \sign(c_3)r }, 
\]
for $k_1,k_2$ given by \eqref{eq:dir}. 
\end{proof}

Finally, we come to the derivation of the asymptotic behavior of the intensity of the maxima in the Fourier transform of the OCT signal for small incident angle.

\begin{proof}[Proof of Lemma~\ref{lem:integral_approximation}]
Since the analysis of the integral in \eqref{eq:int_approx} proceeds along the same lines for $g_{\sigma_\delta,+}$ and $g_{\sigma_\delta,-}$, we simply write $g_{\sigma_\delta}=g_{\sigma_\delta,\pm}$ to make the notation easier.

We assume that locally around $\zeta_0=0,$ we can write $g_{\sigma_\delta}$ as its Taylor series
\[
g_{\sigma_\delta}(\zeta)=\sum_{j\geq 0}\frac{g^{(j)}_{\sigma_\delta}(0)}{j!}\zeta^j.
\]
Using this in \eqref{eq:int_approx}, gives 
\[
\int_\R u_{\sigma_{\bar k},\sigma}(\zeta)g_{\sigma_\delta}(\zeta) d\zeta = \sum_{j\geq 0}\frac{g^{(j)}_{\sigma_\delta}(0)}{j!} \int_\R u_{\sigma_{\bar k},\sigma}(\zeta)\zeta^j d\zeta,
\]
for $j\geq 0.$ We leave out the factor $\frac{1}{2\sqrt{2}}e^{-\tfrac{1}{Q^2}}$ for a moment and calculate this integral for the different values of $j$:
\begin{itemize}
\item
For $j=0$, this leads to the integral 
\[
\int_\R \left(\frac{1}{\sigma^3}-\frac{\zeta^2}{2\sigma^5}\right)e^{-\left(\frac{\zeta}{2\sigma} - i \frac{1}{Q}\right)^2}d\zeta.
\]
After a change of variables $y=\zeta/\sigma,\ \sigma d y= d \zeta$ we obtain
\[
\frac{1}{\sigma^2} \int_\R e^{-\left(\frac{y}{2} - i \frac{1}{Q}\right)^2} d y- \frac{1}{2\sigma^2} \int_\R y^2  e^{-\left(\frac{y}{2} - i \frac{1}{Q}\right)^2} d y =\frac{4\sqrt{\pi}}{\sigma^2Q^2}.  \]
\item
For $j=1,$ we find in the same way
\begin{multline*}
\frac{1}{\sigma} \int_\R y e^{-\left(\frac{y}{2} - i \frac{1}{Q}\right)^2} d y- \frac{1}{2 \sigma} \int_\R y^3  e^{-\left(\frac{y}{2} - i \frac{1}{Q}\right)^2} d y \\
=-4i\sqrt{\pi}\frac{1}{\sigma Q} - \frac{i\sqrt{\pi}}{\sigma} \left(8\frac{1}{Q^3}-12\frac{1}{Q}\right) = \frac{8i\sqrt{\pi}}{\sigma} \left(\frac{1}{Q}-\frac{1}{Q^3}\right).  
\end{multline*}
\item
Similarly, for $j=2,$ we get
\[
\int_\R \left(\frac{\zeta^2}{\sigma^3}-\frac{\zeta^4}{2\sigma^5}\right)e^{-\left(\frac{\zeta}{2\sigma} - i \frac{1}{Q}\right)^2} d\zeta = -\frac{16\sqrt{\pi}}{Q^4} + \frac{40\sqrt{\pi}}{Q^2} + 4(\sqrt{\pi}-3), 
\]
which is constant with respect to $\sigma.$
\item
Following the same procedure, the remaining integrals for $j\geq 3$ are of the form 
\[
\int_\R \left(\frac{\zeta^j}{\sigma^3}-\frac{\zeta^{2+j}}{2\sigma^5}\right)e^{-\left(\frac{\zeta}{2\sigma} - i \frac{1}{Q}\right)^2} d\zeta \simeq C(j) \sigma^{j-2}, 
\]
for a given pre-factor $C(j).$ 
\end{itemize}

The assumption that $\sigma$ is small, let us say $\sigma\ll 1$, yields that the terms of order $\sigma^{-2}$ dominate. Keeping these terms only, results in 
\[
\int_\R f_{\sigma_{\bar k},\sigma}(\zeta)g_{\sigma_\delta}(\zeta) d \zeta \simeq \frac{1}{2\sqrt{2}}e^{-\tfrac{1}{Q^2}} \frac{4\sqrt{\pi}}{\sigma^2 Q^2} g_{\sigma_\delta}(0). 
\]
\end{proof}

\begin{proof}[Proof of Lemma~\ref{lem:maximum}]
We rewrite $F,$ given by  \eqref{eq:sinc_app_shift}, as 
\[
F(\Theta_0; \kappa') = \sinc(\delta\kappa')^2 + \left( \frac{\sin(\delta(\kappa' + 2\Theta_0))}{(\kappa' + 2\Theta_0)}\right)^2 +2\sinc(\delta\kappa') \frac{\sin(\delta(\kappa' + 2\Theta_0))\cos(2 \bar k\Theta_0)}{(\kappa' + 2\Theta_0)}. 
\]
Then, it is clear that 
\[
\lim_{\Theta_0\to \infty} F(\Theta_0; \kappa') = \sinc(\delta\kappa')^2,
\]
which attains its maximum at $\kappa' = 0.$ 
\end{proof}

\section*{Acknowledgements}
This research was supported by the Austrian Science Fund (FWF) in the projects F6803-N36 and F6804-N36 within the Special Research Programme SFB F68:  $``$Tomography Across the Scales$".$

\section*{References}
\printbibliography[heading=none]

\end{document}